\documentclass[aps,prr,twocolumn,superscriptaddress]{revtex4-1}

\usepackage{booktabs}  
\usepackage{graphicx,graphics,xcolor,braket,amsmath,mathtools,subfigure,bm,bbm,tensor,array}
\usepackage{hyperref}
\usepackage{cleveref}
\usepackage{stmaryrd}
\usepackage{amsfonts}
\usepackage{xfrac} 

\crefname{equation}{Eq.}{Eqs.}%
\Crefname{equation}{Equation}{Equations}
\crefname{figure}{Fig.}{Figs.}
\Crefname{figure}{Figure}{Figures}
\crefname{section}{Sec.}{Secs.}
\crefname{subsection}{Sec.}{Secs.}
\Crefname{section}{Section}{Sections}
\crefname{appendix}{Appendix}{Appendices}
\Crefname{appendix}{Appendix}{Appendices}

\usepackage{dsfont}
\usepackage{paralist}

\usepackage{xr-hyper}
\usepackage{multirow}
\usepackage{booktabs,tabularx}

\newcolumntype{R}{>{\raggedleft\arraybackslash\bfseries}X}

\usepackage[normalem]{ulem}

\newcommand{\eqn}[1]{Eq.~(\ref{#1})}

\newcommand{\spect}{\mathsf{Spect}}
\newcommand{\reals}{\mathbb{R}}
\newcommand{\integers}{\mathbb{Z}}

\newcommand{\nicehalfs}{\text{\sfrac{1}{2}}}
\newcommand{\thold}{t_{\rm hold}}

\newcolumntype{P}[1]{>{\centering\arraybackslash}m{#1}}

\begin{document}

\title{
Building a bigger Hilbert space for superconducting devices, one Bloch state at a time
}

\author{Dat Thanh Le} 
\affiliation{ARC Centre for Engineered Quantum System, School of Mathematics and Physics, University of Queensland, Brisbane, QLD 4072, Australia}
\author{Jared~H.~Cole}
\affiliation{Chemical and Quantum Physics, School of Science, RMIT University, Melbourne VIC 3001, Australia}
%
%
%
%

\author{T. M. Stace} \email[]{stace@physics.uq.edu.au}
\affiliation{ARC Centre for Engineered Quantum System, School of Mathematics and Physics, University of Queensland, Brisbane, QLD 4072, Australia}

%
%

\begin{abstract}
Superconducting circuits for quantum information processing are often described theoretically in terms of a discrete charge, or equivalently, a compact phase/flux, at each node in the circuit. 
Here we revisit the consequences of lifting this assumption for  transmon and Cooper-pair box circuits, which are constituted from a Josephson junction and a capacitor, treating both the superconducting phase and charge as non-compact variables.  The periodic Josephson potential gives rise to a Bloch band structure, characterised by the Bloch quasicharge.  We analyse the possibility of creating superpositions of different quasicharge states by transiently shunting inductive elements across the circuit, and suggest a choice of eigenstates in the lowest Bloch band of the spectrum that may support an inherently robust qubit encoding.
\end{abstract}

\maketitle

\section{Introduction}

In 1985, \citet{likharev1985theory} addressed the question of whether the superconducting phase at a circuit node, \mbox{$\hat \phi \!= \! 2\pi \hat{\Phi}/\Phi_0$}, is a compact variable, $\phi\in(-\pi,\pi]$, or whether it is non-compact, $\phi\in\reals$?  These alternatives represent distinct physics: if the phase  is compact, then the spectrum is discrete, while if it is not compact, the spectrum can be a continuum.  For some specific applications, these differences are not material, but  they do lead to different physical predictions in general.  In this paper, we revisit this question and analyse the experimental consequences.


To illustrate the physical differences between the choice of compact or non-compact flux/phase, consider a circuit node within a (quantised) electronic device.  The  flux associated to the node is $\Phi$, which we nondimensionalise using the flux quantum, $\Phi_0=h/(2e)$, to define the superconducting phase \mbox{$\phi=2\pi\Phi/\Phi_0$}.  A conjugate charge, $Q$, is associated to the node, and defining the dimensionless charge $n=Q/(2e)$, we impose the canonical commutation relation $[\hat \phi,\hat n]=i$.  

If  the Hamiltonian for the device is invariant under a subset of phase translations (e.g.\ the $2\pi$-periodic phase-translation symmetry of a  Josephson junction), then it is reasonably common to identify the phase with that of rotor \cite{Koch07}, so that it is compact, and the states \mbox{$\ket{\phi}_{\!\bar \phi}$ and $\ket{\phi+2\pi}_{\!\bar \phi}$} are identical {(the subscript $\bar \phi$ denotes the phase basis)}.   It follows that the Hilbert space in the charge representation is discrete.

The converse holds: if the charge operator is assumed to be discrete, then it follows that the phase is compact.   For instance, \citet{Bouchiat98}
write the (dimensionless) charge operator as
\begin{equation}
\hat n={\sum}_{n\in\mathbb{Z}} n{\ket{n}}_{\!\bar n\!}{\bra{n}},\label{eqn:n}
\end{equation}
where the subscript $\bar n$ indicates the charge basis. 
In this representation, the charge is explicitly quantised: the nondegenerate spectrum of the charge operator is $\integers$.
It follows that phase eigenstates are compact, since 
\begin{equation}
\ket{\phi}_{\!\bar \phi}\equiv{\sum}_{n\in\mathbb{Z}} e^{i n \phi}\ket{n}_{\!\bar n}=\ket{\phi+2\pi}_{\!\bar \phi}.
\end{equation}
That is, $\spect(\hat\phi)=(-\pi,\pi]\Leftrightarrow\spect(\hat n)=\integers$, where $\spect(\hat \bullet)$ denotes the spectrum of the operator $\hat \bullet$.  (In the rest of this paper, we  use the convention that the symbol $\varphi$ denotes a compact phase variable, and the symbol $\phi$ denotes a non-compact phase variable.)

An important corollary  is that the discrete representation of the  charge operator in \cref{eqn:n} prohibits analysis of states with $\phi\notin(-\pi,\pi]$.       This is important if a non-phase-translation invariant device (such as an inductor) is attached to the node.  In this case,  it is typical to assert that $\phi$ is non-compact.  For instance, a conventional model for an oscillator (e.g.\ an $LC$ circuit, or  fluxonium) treats the phase $\phi$ as a non-compact, continuous variable, in which  $\ket{\phi}_{\bar \phi}$ and $\ket{\phi+2\pi}_{\bar \phi}$ are orthogonal  \cite{PhysRevB.60.15398,koch09,Brooks13}.  Physically, this amounts to the observation that changing the magnetic flux of the circuit by $\Phi_0$ yields a new, physically distinct state of the system.   In the phase basis,  the charge operator becomes the differential operator ${}_{\bar \phi\!}\bra{\phi}\hat n\ket{\psi}=-i \partial_\phi \psi(\phi)$, and it follows that \mbox{$\spect(\hat\phi)=\spect(\hat n)=\reals$}.

There are thus two different Hilbert spaces that have been used to describe a circuit node: one that is compact in a canonical variable (phase) and discrete in its conjugate (charge), and another that is non-compact in both conjugate variables. 
 \citet{likharev1985theory} compared the  distinction between these inequivalent descriptions   to  that between a $2\pi$-periodic pendulum and a particle in an extended periodic cosine potential.  Formally,  the Hamiltonians for the two cases appear identical. However the different Hilbert spaces give different spectra.  The spectrum of the pendulum is discrete.  In contrast,  the spectrum of a particle in the periodic lattice consists of continuous Bloch energy bands   \cite{likharev1985theory,Kittel2004}.  The Bloch bands include the pendulum spectrum as a discrete subset, corresponding to  the Bloch eigenvalues at the centre of the Brillouin zone (i.e.\ the $\Gamma$ point, in band-structure nomenclature). 
  It follows that the non-compact periodic system includes the pendulum as a subspace, but its larger Hilbert space admits richer physics.

The difference between the Hilbert spaces in these two descriptions gives rise to a challenge in describing the addition of an inductive element, with Hamiltonian $E_L \hat \phi^2$ to a transmon or Cooper-pair box (CPB) circuit, whose Hamiltonian is \mbox{$E_C \hat n^2-E_J\cos(\hat \phi)$} \cite{koch09}. 
If the  discrete-charge Hilbert space is assumed when $E_L=0$, then the transition to   $E_L>0$ is beset by a catastrophic growth in  the cardinality  of the Hilbert space dimension from $|\integers|$ to $|\reals|$.  
As we discuss later, the dimensional explosion is averted by starting with the assumption that \mbox{$\spect(\hat\phi)=\spect(\hat n)=\reals$} regardless of whether there is an inductor present or not.  In other words, we will adopt the view in Refs.\ \cite{likharev1985theory,koch09} that the superconducting phase and the charge are  inherently non-compact operators.

A related mathematical issue  that arises in the transition from  $E_L=0$ to $E_L>0$ is that the problem is  an instance of singular perturbation theory: the Bloch eigenbands for $E_L=0$, cannot be adiabatically connected to the discrete eigenspectrum for small but nonzero $E_L$ \cite{koch09}.  We discuss this in detail in \cref{sec:ind}.  

In what follows, we review a convenient basis proposed by \citet{Zak67}, and use this to reanalyse the circuit for a transmon/CPB.  With a non-compact phase, a $2\pi$ phase-periodic device will have an  eigensystem arranged into Bloch bands, which are labelled by a band index and a ``quasicharge".  In the lowest Bloch band, eigenstates are comb-like periodic functions of the phase, which is reminiscent of the oscillator code states proposed in \citet{Gottesman01}.  In \cref{sec:symmetrybreaking} we describe experiments that would break the $2\pi$ periodicity in order to controllably create and observe superpositions of the Bloch quasicharge states in the lowest band.  Section \ref{sec:discussion} then addresses the properties of such superpositions, which includes inherent robustness against relaxation, and possible experimental considerations.

\section{Revisiting transmon/Cooper-pair box in the Zak basis}

\subsection{The Modular Zak Basis}

The Zak basis   \cite{Zak67,le2019doubly} is useful and interesting because it provides a compact 2D representation of a Hilbert space that is non-compact in a 1D coordinate (e.g.\ the position or phase, or their conjugate variables).  The basis set, $\{\ket{k,\varphi}\}$, is labelled by the modular \footnote{\emph{modular} is used to indicate that the labels for the basis states are in the set $k\in\reals/(\integers-1/2)=(-\nicehalfs,\nicehalfs]$ for the modular charge and $\varphi\in\reals/(2\pi \integers-\pi)=(-\pi,\pi]$ for the modular flux.} charge $k\in(-\nicehalfs,\nicehalfs]$ and modular flux $\varphi\in(-\pi,\pi]$. We will show that this resolves some of the technical difficulties when a large inductance (i.e.\ small $E_L$) is shunted across a transmon/CPB circuit.  Here we give a brief review of this basis.

The Zak basis states can be defined  with respect to either the non-compact phase basis, denoted as $\bar\phi$, or the non-compact charge basis, $\bar n$, as  
\begin{align}
    \left|k,\varphi \right\rangle &=  \sum\limits_{j = -\infty}^{\infty} \! e^{i 2\pi j k }\!\ket{\varphi - 2\pi j }_{\!\bar\phi}\label{phasebasis}\\
   & =\tfrac{e^{i k \varphi}}{\sqrt{2\pi}}\!\sum\limits_{j=-\infty}^{\infty} \!e^{-i j \varphi } \!\ket{ j-k }_{\!\bar n},     \label{chargebasis}
\end{align}
with $k\in(-\nicehalfs,\nicehalfs]$ and  $\varphi\in(-\pi,\pi]$. 
These states satisfy orthonormalisation \mbox{$\langle k,\varphi\ket{k'\!,\varphi'}=\delta(k-k')\delta(\varphi\!-\!\varphi')$}, {and}
the generalised periodic boundary identities
\mbox{$ \ket{ -1/2, \varphi} \! =\! \ket{ 1/2, \varphi }$} and \mbox{$\ket{ k, - \pi} \!=\! e^{-i 2\pi k} \!\ket{ k, \pi}$}.

The Zak basis is convenient for several reasons.  Firstly, it is compact in the 2D space of $k$ and $\varphi$, which will turn out to be helpful for numerical calculations.

Secondly, it is complete: using \cref{phasebasis}, it is straightforward to show that 
\begin{equation}
\int_{-\pi}^\pi \!d\varphi \int_{-1/2}^{1/2} \!dk\,{\ket{k,\varphi}}{\bra{k,\varphi}}=\int_{-\infty}^\infty d\phi  {\ket{\phi}}_{\!\bar\phi}{\bra{\phi}} =\mathds{1}.
\end{equation}
A corollary is that two phase states which are offset from one another by an integer multiple of $2\pi$ have orthogonal representations in this basis.  


Thirdly, polynomial and (some) periodic functions of charge and flux have local representations in the Zak basis.  In particular, we have \cite{Ganeshan16}
\begin{align}
\bra{k,\varphi} \hat{n} \ket{\psi} &=  - i \tfrac{\partial }{\partial \varphi} \psi(k,\varphi),\label{eqnn} \\
\bra{k,\varphi} \hat{\phi} \ket{\psi} &= \big( - i \tfrac{\partial }{\partial k} + \varphi \big) \psi(k,\varphi),\label{eqnphi}
\end{align}
and 
\begin{eqnarray}
\bra{k,\varphi} \cos(2\pi\hat{n}) \ket{\psi} &=&  \cos(2\pi k)  \psi(k,\varphi), \\
\bra{k,\varphi} \cos(\hat{\phi}) \ket{\psi} &=& \cos(\varphi) \psi(k,\varphi),
\end{eqnarray}
where $\psi(k,\varphi)\equiv\langle k,\varphi | \psi \rangle$.

This is in contrast to the flux or charge bases, for which periodic functions have a delocalised representation, e.g. 
\begin{eqnarray}
\cos(\hat\phi)\!&=&\!\tfrac{1}{2}(e^{i\hat\phi}+e^{-i\hat\phi})\nonumber\\
&=&\!\tfrac{1}{2}\int_{-\infty}^{\infty} dq\big( {\ket{q+1}}_{\!\bar n\!}{\bra{q}}+{\ket{q}}_{\!\bar n\!}{\bra{q+1}}\big),
\end{eqnarray}
 is delocalised in the charge basis.

\begin{figure}
\begin{center}
\includegraphics[width=0.9\columnwidth]{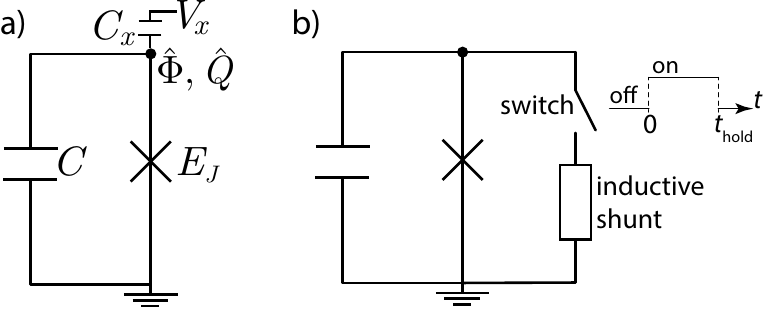}
\caption{({\sf a}) We use the Zak basis to analyse the circuit for a transmon or Cooper-pair box (CPB), consisting of a Josephson junction and a capacitor, which is characterised by a flux $\hat \Phi$ and a conjugate charge $\hat Q$.  An external voltage source, $V_x$, may be capacitively coupled to the node of the circuit. ({\sf b})  An inductive element (box) is transiently switched into the circuit for the time interval $0<t<\thold$.  The inductive element may be a linear inductor, which we analyse in \cref{sec:ind} or a non-linear $4\pi$-periodic element, which we analyse in \cref{sec:4pi}.} \label{fig:transmon}
\end{center}
\end{figure}

\subsection{Band Structure of the Transmon/Cooper-pair box Hamiltonian}

Having introduced a convenient basis that has a local representation of both polynomial and periodic functions of $\hat n$ and $\hat \phi$, we now revisit the well-studied  Hamiltonian  describing  CPBs and transmons, depending on the parameter choices.

The results of this section  can be derived straightforwardly in the phase basis using Bloch's theorem.  However, the Zak basis is helpful for describing elements that break the dynamical $2\pi$-phase periodicity in the Hamiltonian,  which we address later.  In this section we therefore re-derive the Bloch wavefunctions in the Zak basis.

We consider the circuit pictured in \cref{fig:transmon}a, consisting of  a Josephson junction and a capacitor, and subject to an external voltage bias $V_x$. This circuit is  described by the Hamiltonian {\cite{Koch07,Spiller1990}}
\begin{equation}
\hat H_{\sf tmon}=E_C (\hat n-n_x)^2-E_J \cos(\hat\phi),\label{TransmonHamiltonian1}
\end{equation}
where $E_C=2e^2/(C+C_x)$ {is the charging energy with $C$ the Josephson junction capacitor and $C_x$ the external capacitor}, $n_x=C_x V_x/(2e)$ is the gate voltage referenced to a dimensionless on-site charge shift, and $E_J$ is the junction energy.  This Hamiltonian describes a CPB when $E_J\ll E_C$ {\cite{Bouchiat98}}, and a transmon when $E_J \gg E_C$ {\cite{Koch07}}.

Representing states of the system in the Zak basis, $\psi(k,\varphi)=\langle k,\varphi\ket{\psi}$, yields the time-independent Schr\"odinger equation
\begin{equation}
E_C(-i   \tfrac{\partial }{\partial \varphi}-n_x)^2\psi(k,\varphi)-E_J\cos(\varphi)\psi(k,\varphi)=E \psi(k,\varphi),\label{TransmonHamiltonian}
\end{equation}
with generalised, periodic boundary conditions 
\begin{align}
 \psi( - 1/2, \varphi ) &= \psi ( 1/2, \varphi), \label{bc1}
\\
 \psi (k, - \pi ) &= e^{2\pi k i} \psi (k, \pi).\label{bc2}
\end{align}
Since the only derivatives in \cref{TransmonHamiltonian} act on $\varphi$, the modular charge $k$ is a constant of motion. 

\begin{figure}
\begin{center}
\includegraphics[width=\columnwidth]{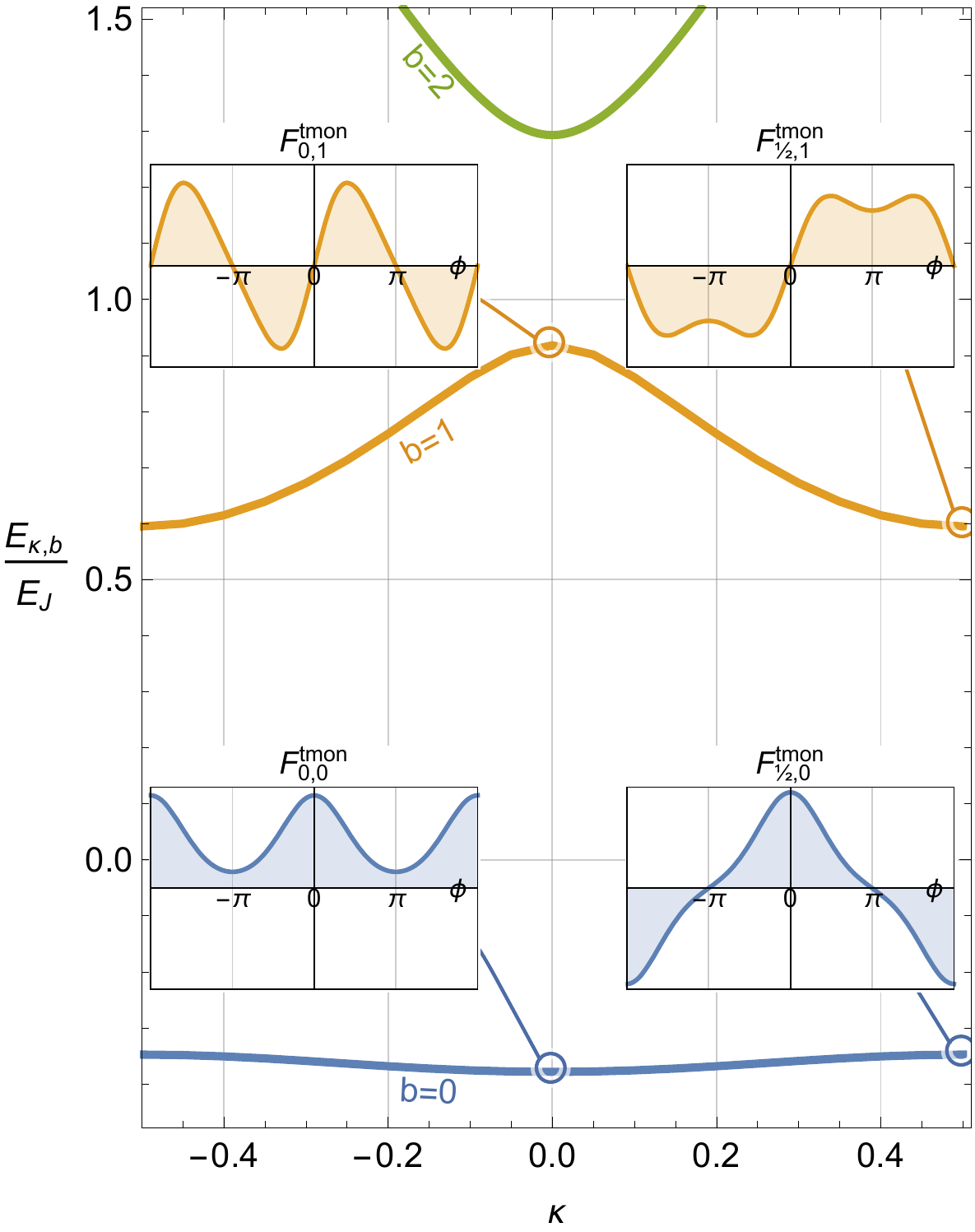}
\caption{The spectrum of the Hamiltonian in \cref{TransmonHamiltonian1}, as a function of the quasicharge, $\kappa$, with $E_C/E_J=1$, and \mbox{$n_x=0$}. {Note that the regime $E_J/E_C=1$ is intermediate \cite{PhysRevX.6.031016} between the transmon and CPB ones and is used here for illustrative purpose only.} Bands are labelled by the band index $b$.  In this regime, the lowest band is quite narrow relative to the gap.  Inset figures depict selected Bloch wavefunctions $F^{\sf tmon}_{\kappa,b}(\phi)$ in the non-compact phase basis, given in \cref{psiF}, for the lowest two bands,  at the centre and at the edge of  the Brillouin zone.  The vertical axis in the insets is arbitrary, but scaled identically for all. 
The restriction of $F^{\sf tmon}_{\kappa,b}$ to the compact domain $(-\pi,\pi]$ yields the form on the RHS of \cref{eqn:deltaF}.  The  inset modes at the centre and edge of the lowest Bloch band are comb-like states, with periodicity $2\pi$ and $4\pi$ respectively. } 
\label{bandstructure}
\end{center}
\end{figure}

We solve \cref{TransmonHamiltonian} using separation of variables. Substituting $\psi (k, \varphi)=K(k)F(\varphi)$  into \cref{TransmonHamiltonian}, we see that $K$ factorises out of the resulting expression, yielding
\begin{equation}
E_C(-i   \tfrac{\partial }{\partial \varphi}-n_x)^2F(\varphi)-E_J\cos(\varphi)F(\varphi)=E \,F(\varphi).\label{FSE}
\end{equation}
The boundary condition \cref{bc2} requires that   
\begin{equation}
F(-\pi)=e^{2\pi k i}F(\pi).\label{eqnbcF}
\end{equation}  Since $F$ is supposed to be independent of $k$, this boundary condition appears to lead to a contradiction in the separability assumption.  However, if $K(k)=\delta(k-\kappa)$ for some constant $\kappa$, then $F$ may depend parametrically on $\kappa$ without violating  separability \footnote{We note that using  this form for $K$  in the separable ansatz for $\psi$ satisfies the boundary condition \cref{bc1}, except at the Brillouin zone boundary, where $\kappa=\pm\nicehalfs$ both label the same state.  This is resolved by symmetrising, i.e.\ taking \mbox{$K(k)=\delta(k-\nicehalfs)+\delta(k+\nicehalfs)$} for this special case.}.  Thus the modular charge is promoted to a quantum number, $k\rightarrow\kappa$, which is commonly called the \emph{Bloch wavenumber}, or alternatively the \emph{quasicharge} in   \citet{likharev1985theory}.  

Thus transmon/CPB eigenstates $\ket{\psi}$ are expressed in the Zak basis as
\mbox{$
\psi (k, \varphi)=\delta(k-\kappa)F_{\kappa}(\varphi)
$},
 where  \mbox{$F_\kappa(-\pi)=e^{2\pi \kappa i}F_\kappa(\pi)$}, which follows from \cref{eqnbcF}. 
This generalised periodic boundary condition implies that $F_\kappa(\varphi)$ are the Bloch wavefunctions.  This is made clear by defining the periodic function \mbox{$u_\kappa(\varphi)=e^{i \kappa \varphi}F_{\kappa}(\varphi)$}, which satisfies the Bloch equation 
\begin{equation}
E_C\big(-i   \tfrac{\partial }{\partial \varphi}-(n_x+\kappa)\big)^2u_\kappa(\varphi)-E_J\cos(\varphi)u_\kappa(\varphi)=E u_\kappa(\varphi)\label{uSE},
\end{equation} 
with $u_\kappa(\varphi)=u_\kappa(\varphi+2\pi)$.  

As a result, the transmon/CPB eigenenergies, $E^{\sf tmon}_{\kappa,b}$, and eigenstates, $\ket{\psi^{\sf tmon}_{\kappa,b}}$, are labelled by the wavenumber  $\kappa\in(-1/2,1/2]$ and the band index $b\in\mathbb{N}$, giving rise to the usual Bloch bands. {Here, for brevity we label the eigensystem with the superscript  ``$\textsf{tmon}$" for ``transmon", but with the understanding that these arguments also apply for CPBs.}
In the compact Zak basis, the eigenfunctions are
\begin{equation}
\psi^{\sf tmon}_{\kappa,b} (k, \varphi)\equiv\langle{k,\varphi}\ket{\psi^{\sf tmon}_{\kappa,b}}=\delta(k-\kappa)F^{\sf tmon}_{\kappa,b}(\varphi).\label{eqn:deltaF}
\end{equation}

It also follows from \cref{uSE}, and Bloch's theorem, that in the non-compact phase basis, the transmon/CPB eigenfunctions are
\begin{equation}
\psi^{\sf tmon}_{\kappa,b}(\phi)\equiv{}_{\!\bar\phi\!}\langle\phi\ket{\psi^{\sf tmon}_{\kappa,b}}=e^{-i \kappa \phi}u_\kappa(\phi)=F^{\sf tmon}_{\kappa,b}(\phi),\label{psiF}
\end{equation}
 where $\phi\in\reals$.  That is, $F$ is the eigenfunction in the non-compact phase basis.

\Cref{bandstructure} depicts the band-structure and selected eigenfunctions  for the Hamiltonian, \cref{TransmonHamiltonian1}, for  \mbox{$E_C/E_J=1$, with $n_x=0$}.  As an aside, we note that $n_x$ just sets the origin of the band-structure, so there is no loss of generality in setting $n_x=0$.  In this  regime, the bands are relatively flat, so that the band-width of the lowest band is narrow  relative to the  band gaps. For these parameters, the transition energy at the centre of the  Brillouin zone is approximately given by the harmonic oscillator energies that would be associated to the linearised system, $\sqrt{2 E_JE_C} $, i.e.\ numerically we find that
\begin{equation}
(E_{\kappa=0,b=1}-E_{0,0})/ \sqrt{2 E_JE_C} \approx0.9,
\end{equation}  is close to unity.

The eigenfunction at the centre of the Brillouin zone, $\psi^{\sf tmon}_{\kappa=0,b}(\phi)$, is $2\pi$-periodic in the non-compact phase basis, 
while the eigenfunction at the edge of the Brillouin zone, $\psi^{\sf tmon}_{\kappa=\nicehalfs,b}(\phi)$, is $4\pi$-periodic.   This is  illustrated in the  insets in \cref{bandstructure}.
In the lowest band, $b=0$, the states at the centre and at the edge of the Brillouin zone are comb-like,  similar to the code states of the oscillator `GKP' code proposed by \citet{Gottesman01}, which have previously been discussed for stable superconducting qubits \cite{Brooks13}.  The width of the peaks in the comb scales as $\sqrt{E_C/E_J}$, which sets the dimensionless  zero-point modular displacement of the device.




%
%

\section{Transient symmetry breaking}\label{sec:symmetrybreaking}

The previous section re-derived standard results from Bloch's theorem for a particle in a periodic potential over a non-compact domain. The Zak basis becomes particularly helpful for analysing the situation of a  transmon or CPB that is transiently shunted by an electronic element that breaks the $2\pi$-periodicity of the Josephson junction, such as a linear inductor or a (phenomenological) nonlinear $4\pi$-periodic  inductive element.  This is illustrated in \cref{fig:transmon}b, showing the transmon/CPB circuit with a switch that allows a linear or nonlinear inductive shunt element (box) to be transiently shunted across the circuit.  This circuit has similarities to the gates proposed for the $0{\text -}\pi$ qubit \cite{kitaev2006protected,Brooks13,Paolo19,Groszkowski2018}, which we touch on later.

In this section we solve the transient evolution for both cases.  Since the  $2\pi$-periodicity of the Josephson junction is broken, the  eigenstates of the transmon/CPB Hamiltonian become coupled, and the evolution of the system generates superpositions of the Bloch eigenstates of the bare  system. 
In particular, we will show that it is possible to  controllably create coherent superpositions of states at the centre and edge of the Brillouin zone in the lowest band, i.e.
\begin{equation}
\ket{\Psi_{b=0}}=\alpha \ket{\psi^{\sf tmon}_{0,0}}+\beta \ket{\psi^{\sf tmon}_{\nicehalfs,0}}.\label{superposition}
\end{equation}
The possibility of producing and measuring such a superposition is the core development in this paper.

\subsection{Transiently shunted linear inductor}\label{sec:ind}

Here we suppose that the inductive shunt is a linear inductor, with inductance $L$.  
  The Hamiltonian describing this system is {\cite{Manucharyan09,koch09}}
\begin{equation}
\hat H=E_C \hat n^2+E_L(t) \hat \phi^2-E_J \cos(\hat\phi),\label{shuntedTransmonHamiltonian1}
\end{equation}
where  
\begin{equation}
E_L(t)=\left\{\begin{array}{cl}
0, &t<0 \\ 
E_L, &0<t<\thold \\
0, &  t>\thold\end{array}\right. ,
\end{equation}
for  $E_L=\Phi_0^2/(8\pi^2L)$, and we set the external voltage bias to zero (i.e.\ $n_x=0$).  For a time-independent inductive term, we note that this Hamiltonian would describe a fluxonium circuit \cite{Manucharyan09,koch09}, up to some choice of flux bias.
 While the switch is on, the inductor explicitly breaks the $2\pi$-periodicity of the junction Hamiltonian, making  it  very clear that the phase $\phi$ is not a compact variable. {In \cref{app:indeigenmodes}, the  fluxonium states in the Zak basis  are plotted and compared to the transiently inductor-shunted transmon/CPB states.}

\begin{figure*}
\begin{center}
\includegraphics[height=3.9cm]{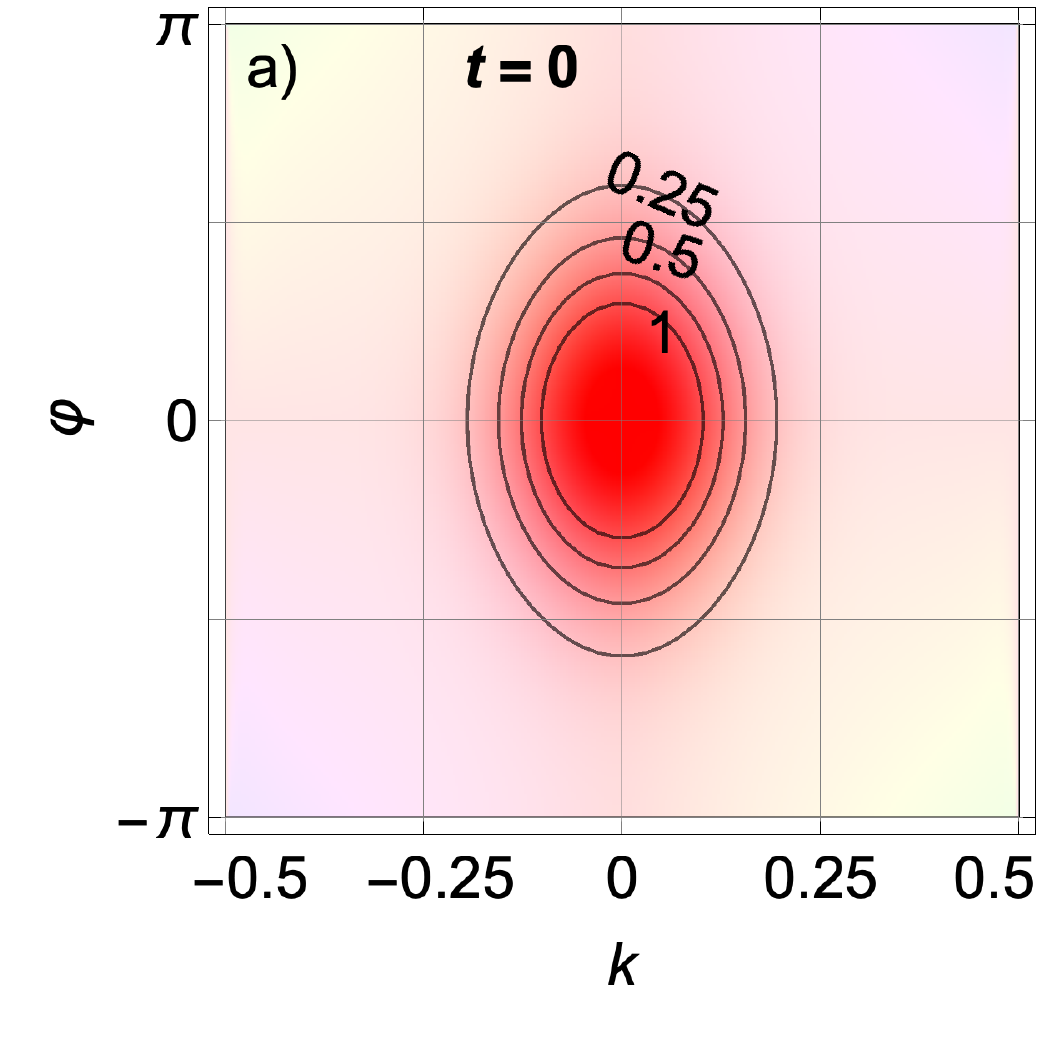}
\includegraphics[height=3.9cm]{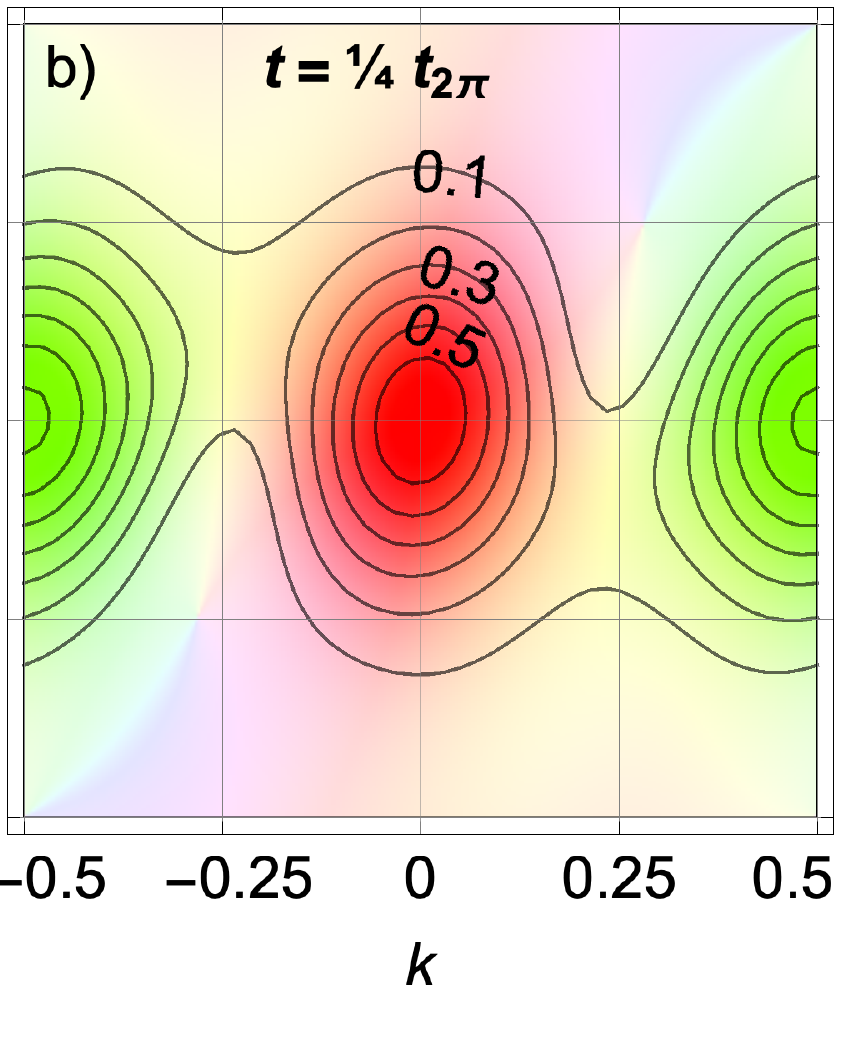}
\includegraphics[height=3.9cm]{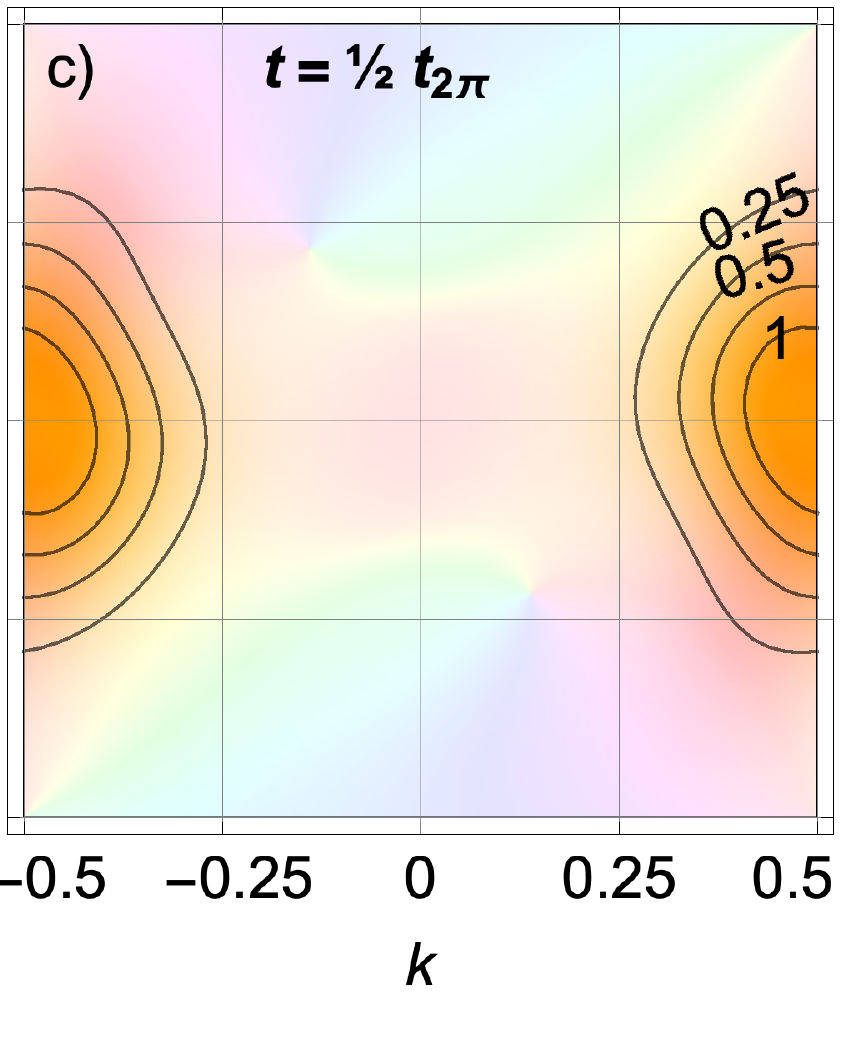}
\includegraphics[height=3.9cm]{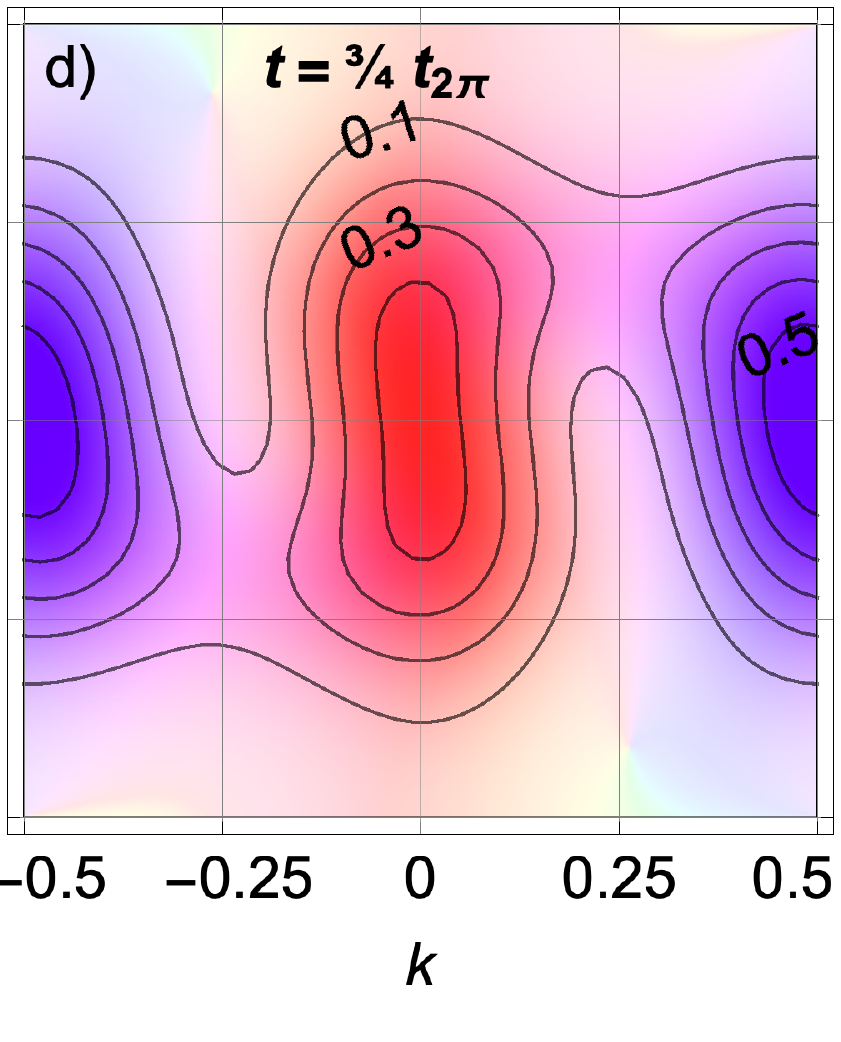}
\includegraphics[height=3.9cm]{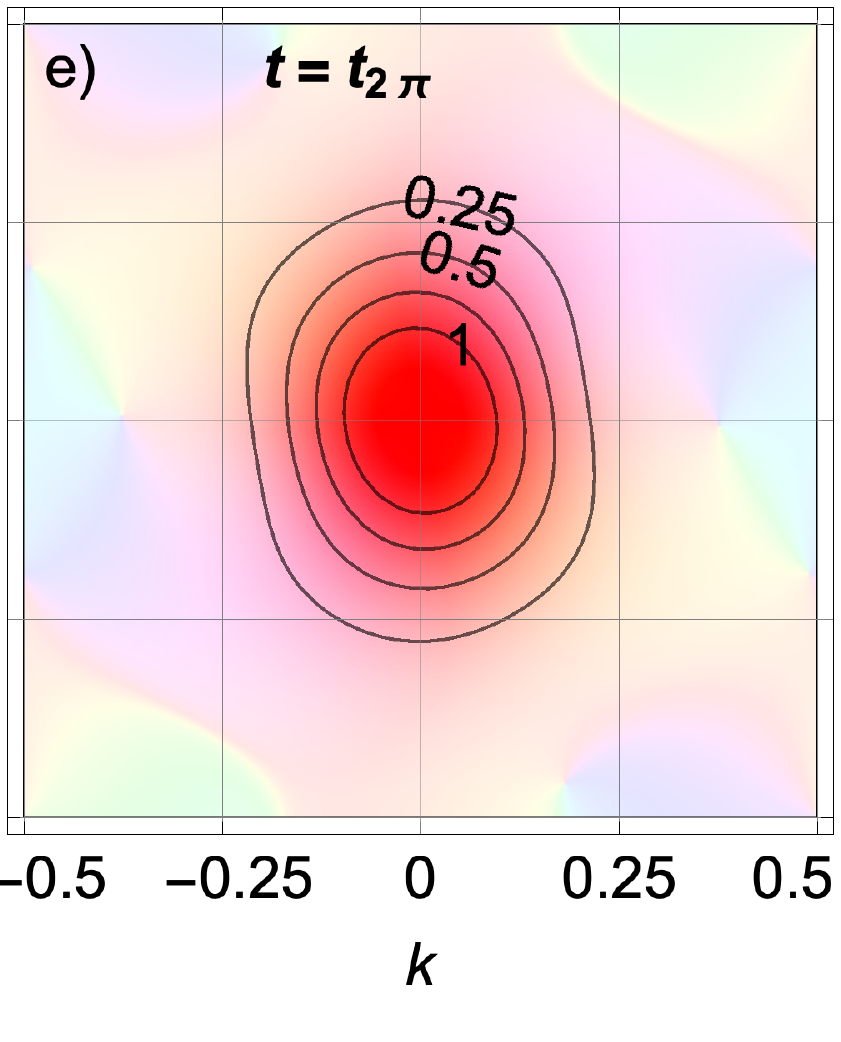}
\hfill\raisebox{0.7cm}{\includegraphics[width=0.9cm]{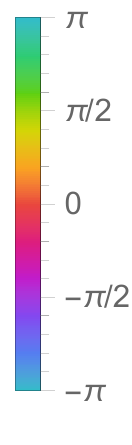}}
\caption{(a -- e) Snapshots of the time evolution of the wavefunction probability density, $|\psi^{\sf ind}(t;k,\varphi)|^2$, of a transmon transiently shunted by an inductor (contours).  The hue shows the local phase of the wavefunction, $\arg(\psi^{\sf ind})$,  relative to the phase at the origin, enumerated  in the scale bar.  The initial state is a (nearly-)separable, $\Delta$-broadened state given in \cref{eqn:indinitialstate}, 
and we have set the (dimensionless) broadening parameter to $\Delta=0.2$.  For these plots, other parameters are $E_C/E_J=1$, $E_L/E_J=1/(2\pi)^2$, and $n_x=0$. For numerics, we have generated the evolution using the first 100 eigenmodes. 
} \label{fig:timeEvolution}
\end{center}
\end{figure*}

The inductive shunt in \cref{shuntedTransmonHamiltonian1} breaks the periodic flux-translation  symmetry, and is unbounded above, so the spectrum of eigenstates, $\{E_{j\in\mathbb{N}}\}$, is discrete. In the Zak basis,  the energy eigenfunctions \mbox{$\langle{k,\varphi}\ket{\psi^{\sf ind}_j}\equiv\psi^{\sf ind}_j(k,\varphi)$} satisfy the time-independent Schr\"odinger equation
\begin{equation}
-E_C  \tfrac{\partial^2 \psi^{\sf ind}_j }{\partial \varphi^2}+E_L \big( \!-\! i \tfrac{\partial }{\partial k} \!+\! \varphi \big)  ^2\!\psi^{\sf ind}_j\!-\!E_J\cos(\varphi)\psi^{\sf ind}_j\!=\!E_j \psi^{\sf ind}_j,\label{shuntedTransmonHamiltonian}
\end{equation}
along with the boundary conditions \cref{bc1,bc2}. Here the superscript ``${\sf ind}$" labels the eigenmodes of the inductively shunted device.

In the rest of this section, we wish to solve the evolution generated by \cref{shuntedTransmonHamiltonian}, assuming the system is initialised in the  transmon/CPB ground state, $\ket{\psi^{\sf tmon}_{0,0}}$, at $t=0$.  In the Zak basis this is given by  \cref{eqn:deltaF}:
\begin{equation}
\psi^{\sf tmon}_{0,0} (k, \varphi)= \delta(k)F^{\sf tmon}_{0,0}(\varphi).\label{initistateind}
\end{equation}
The  ground state Bloch wavefunction, $F^{\sf tmon}_{0,0}$, is shown as an inset in \cref{bandstructure}.

Unfortunately, \cref{shuntedTransmonHamiltonian} presents a technical challenge in dealing with situations where $E_L$ is small.    In particular, $E_L$ appears as a coefficient on the highest order derivative with respect to $k$, so the limit $E_L\rightarrow 0^+$ is an instance of singular perturbation theory.  
Singular perturbation problems  are often solved by introducing a small `length' scale, over which the singular perturbation relaxes \footnote{Since the order of a PDE determines the number of boundary conditions that need to be specified, taking $E_L=0$ results in a lower order PDE than any value $E_L\rightarrow0^+$. Archetypal singular value problems occur in the Navier-Stokes equation  for laminar flow of viscous fluids, where the small parameter is the inverse of the Reynolds number, and the relaxation length scale corresponds to the boundary layer thickness.}.  Inside the small length scale, the singular perturbation is significant, outside this scale, it becomes negligible.  

In our case, the singular term in \cref{shuntedTransmonHamiltonian} corresponds to a diffusive operator, $-E_L\partial^2_k$, in the Schr\"odinger equation with a small  coefficient. If  $E_L=0$, then the spectrum of the Hamiltonian is the continuum band-structure discussed in the previous section, and shown in \cref{bandstructure}.  However if $E_L>0$ is arbitrarily small but nonzero, then the spectrum becomes discrete, so that there is no adiabatic way to connect the spectra as $E_L$ changes from 0.  In practice, this means that systems  which are initially highly localised in the $k$ coordinate will diffuse rapidly over very short times, but this coherent  diffusion becomes insignificant as the wavefunction spreads out.  Physically this corresponds to large, transient currents through the shunting inductor as the modular charge redistributes across the device.

As such we now solve the evolution generated by \cref{shuntedTransmonHamiltonian} as an initial value problem, and regularise the rapid diffusion at short times (arising from the singular perturbation) by replacing the $\delta$ function in  \cref{initistateind} with a $\Delta$-broadened gaussian:
\begin{equation}
\delta(k)\rightarrow\delta_\Delta(k,\varphi) \equiv \mathcal{N} e^{-ik\varphi}(e^{-(k/\Delta)^2}-e^{-(\nicehalfs/\Delta)^2}).\label{eqn:broaddelta}
\end{equation}Here  $ \mathcal{N}$ 
 is a normalisation chosen to satisfy $\int_{-\nicehalfs}^\nicehalfs dk |\delta_\Delta(k,\varphi)|^2=1$, and we introduce 
 the  offset $e^{-(\nicehalfs/\Delta)^2}$ and complex phase $e^{-ik\varphi}$ on the RHS of \cref{eqn:broaddelta} to ensure that  $\delta_\Delta$ satisfies the boundary conditions \cref{bc1,bc2} respectively.  

We then (implicitly) define broadened transmon eigenmodes $\ket{\psi^{\sf tmon}_{\kappa,b}}_{\!\Delta}$  through
\begin{equation}
\langle{k,\varphi}\ket{\psi^{\sf tmon}_{\kappa,b}}_{\!\Delta} = \delta_\Delta(k,\varphi)F^{\sf tmon}_{\kappa,b}(\varphi).\label{eqn:Deltabroadened}
\end{equation}
This allows us to define  the  $\Delta$-broadened ground state of the transmon with which we initialise the system,
\begin{equation}
\ket{\psi^{\sf ind} (t=0)}= \ket{\psi^{\sf tmon}_{0,0}}_{\!\Delta}.\label{eqn:indinitialstate}
\end{equation}

For the rest of this section, we use the parameter values $E_C/E_J=1$,  $E_L/E_J=1/(2\pi)^2\ll1$, and $n_x=0$. To calculate the time evolution, we decompose the initial state into the  eigenmodes  of the inductively-shunted transmon {(i.e. the fluxonium)}.  For reference, some eigenmodes and eigenenergies are plotted in the Zak basis in  \cref{app:indeigenmodes}.  In spite of the smallness of $E_L$, the spectrum is discrete, emphasising the effect of the singular perturbation relative to the continuum band-structure when $E_L=0$.
  
Snapshots of the time-evolved probability density $|\psi^{\sf ind}(t;k,\varphi)|^2$ are shown as contours in \cref{fig:timeEvolution}.  The initial state is localised around $k=0$, and somewhat delocalised in $\varphi$ (for visualisation, we choose $\Delta=0.2$).  At a later time, $t_{2\pi}=2\pi/(E^{\sf ind}_1-E^{\sf ind}_0)$, the system returns to a state that is localised around $k=0$, and has a large overlap with the initial state.  For the parameter choices in \cref{fig:timeEvolution}, $t_{2\pi}\approx 6.8/E_J$.  At intermediate times, the state is delocalised over $k$. In particular, at  $t_{2\pi}/4$ and $3t_{2\pi}/4$, the state is a superposition of approximately-localised modes with  support around \mbox{$k=0$} and  $k=\nicehalfs$, i.e.\ it is approximately \mbox{$(\ket{\psi^{\sf tmon}_{0,0}}_{\!\Delta}+e^{i\theta}\ket{\psi^{\sf tmon}_{\nicehalfs,0}}_{\!\Delta})/\sqrt{2}$}.  
At time $t_{2\pi}/2$, the state is localised around $k=\nicehalfs$.

\begin{figure}
\begin{center}
\includegraphics[width=8cm]{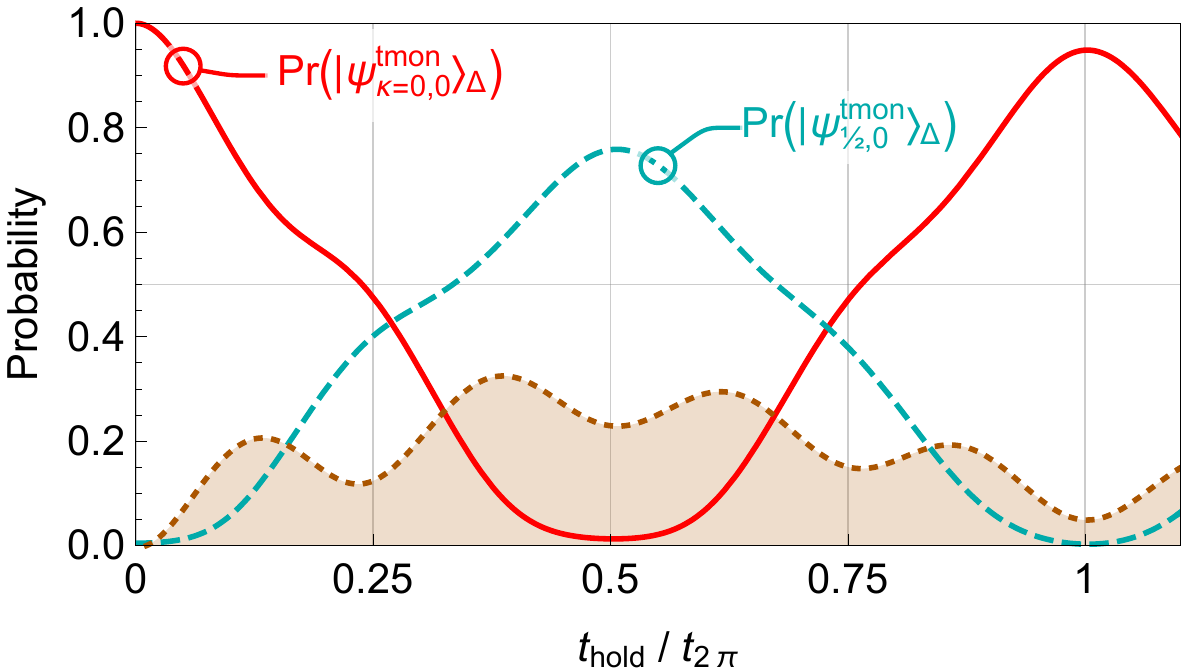}
\caption{Time-evolution of the return probability of the inductively-shunted transmon (solid-red), initialised in the $\Delta$-broadened state $ \ket{\psi^{\sf tmon}_{0,0}}_{\!\Delta}$, \cref{eqn:indinitialstate}.  Also shown is the probability of the  $\Delta$-broadened state localised near the Brillouin zone boundary, $ \ket{\psi^{\sf tmon}_{\nicehalfs,0}}_{\!\Delta}$ (dashed-blue), along with the residual probability in other states (dotted-shaded).  The parameter values used in the simulation are $E_C/E_J=1$, $E_L/E_J=1/(2\pi)^2$ and $\Delta=0.2$.} \label{indoverlap}
\end{center}
\end{figure}


This evolution is confirmed by computing the return probability, $|{}_{\Delta}\!\langle{\psi^{\sf tmon}_{0,0}}\ket{\psi^{\sf ind}(t)}|^2$, as a function of the switch hold time, $\thold$, as well as the time-dependent probability for the system to be in the state $\ket{\psi^{\sf tmon}_{\nicehalfs,0}}_{\!\Delta}$, i.e.\ $|{}_{\Delta}\!\langle{\psi^{\sf tmon}_{\nicehalfs,0}}\ket{\psi^{\sf ind}(t)}|^2$.  These quantities are plotted in \cref{indoverlap}, confirming oscillations between the initial state near $k=0$  and a state near the edge of the Brillouin zone.  We note that  the incomplete overlap with $\ket{\psi^{\sf tmon}_{\nicehalfs,0}}_{\!\Delta}$ at $t_{2\pi}/2$ is a consequence of the anharmonic Schr\"odinger evolution for this nonlinear system which mildly distorts the localised wavepackets at intermediate times -- since the evolution is unitary, the state remains pure at all times.  The dotted-shaded curve shows the residual probability to be in a state other than these two modes. {Specifically, the ideal state $\ket{\psi^{\sf tmon}_{\nicehalfs,0}}_{\Delta}$, similar to $\ket{\psi^{\sf tmon}_{0,0}}_{\Delta}$,  has a probability distribution $\vert \langle{k,\varphi}\ket{\psi^{\sf tmon}_{\nicehalfs,0}}_{\!\Delta} \vert^2$ that is symmetric about $\varphi = 0$ and $k=0$. However, in the middle snapshot of \cref{fig:timeEvolution} there is a small asymmetry in the distribution, so the overlap with $\ket{\psi_{\nicehalfs,0}^{\sf tmon}}_{\Delta}$ is not perfect.  }

This result shows that a transmon which is transiently shunted by an inductor will become delocalised over the modular charge, $k$.  If the inductor is switched out of the circuit at $\thold=t_{2\pi}/4$ or $3t_{2\pi}/4$, the device will be left in a superposition of different transmon/CPB eigenstates which approximate the  state given in \cref{superposition}.

\subsection{Transiently shunted  $4\pi$-periodic device}\label{sec:4pi}

An alternative approach to coupling different $k$ states is to  break the $2\pi$-periodic $\phi$-translation symmetry by shunting the transmon with a $4\pi$-periodic element \cite{PhysRevB.77.184506,PhysRevB.79.161408}.  
For the purposes of this section, we do not describe the implementation of such an element \cite{Ginossar:2014aa,PhysRevX.6.031016,Laroche:2019aa,Mota19}, but begin with a phenomenological Hamiltonian {\cite{Mota19}}
 \begin{equation}
\hat H=E_C \hat n^2-E_J \cos(\hat\phi)-E_{\sf 4\pi}(t) \cos(\hat\phi/2),\label{shuntedTransmonHamiltonian4pi}
\end{equation}
where
\begin{equation}
E_{\sf 4\pi}(t)=\left\{\begin{array}{cl}
0, &t<0 \\ 
E_{\sf 4\pi}, &0<t<\thold \\
0, &  t>\thold\end{array}\right. .\label{e4pi}
\end{equation}
We note in passing that switching control could be established in this system by forming a SQUID-like configuration of parallel $4\pi$-periodic elements with a flux bias, so that the $E_{\sf 4\pi}$ may be controllably tuned by an external bias current, in analogy with the switchable junction in \citet{Brooks13}.

Along with the transient control in \eqn{e4pi},  
we assume the system is initialised in the transmon ground state, 
\begin{equation}
\ket{\psi^{\sf 4\pi}(t=0)}=\ket{\psi^{\sf tmon}_{0,0}},\label{4piinit}
\end{equation}
which is localised at \mbox{$k=0$}, as in \cref{initistateind}. 
Since the Hamiltonian term associated to such a device is bounded, the problems arising from singular perturbations  are avoided, and there is no need to invoke $\Delta$ broadened initial states.

\begin{figure}[t]
\begin{center}
\includegraphics[width=8cm]{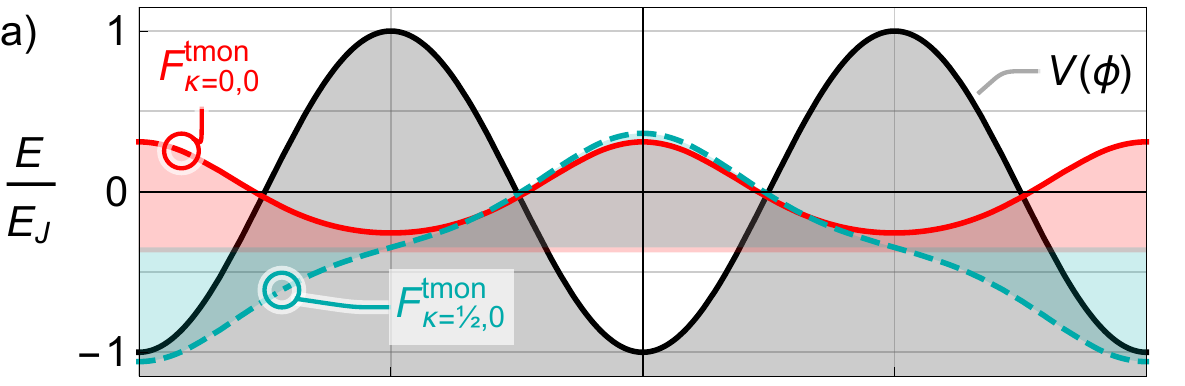}
\includegraphics[width=8cm]{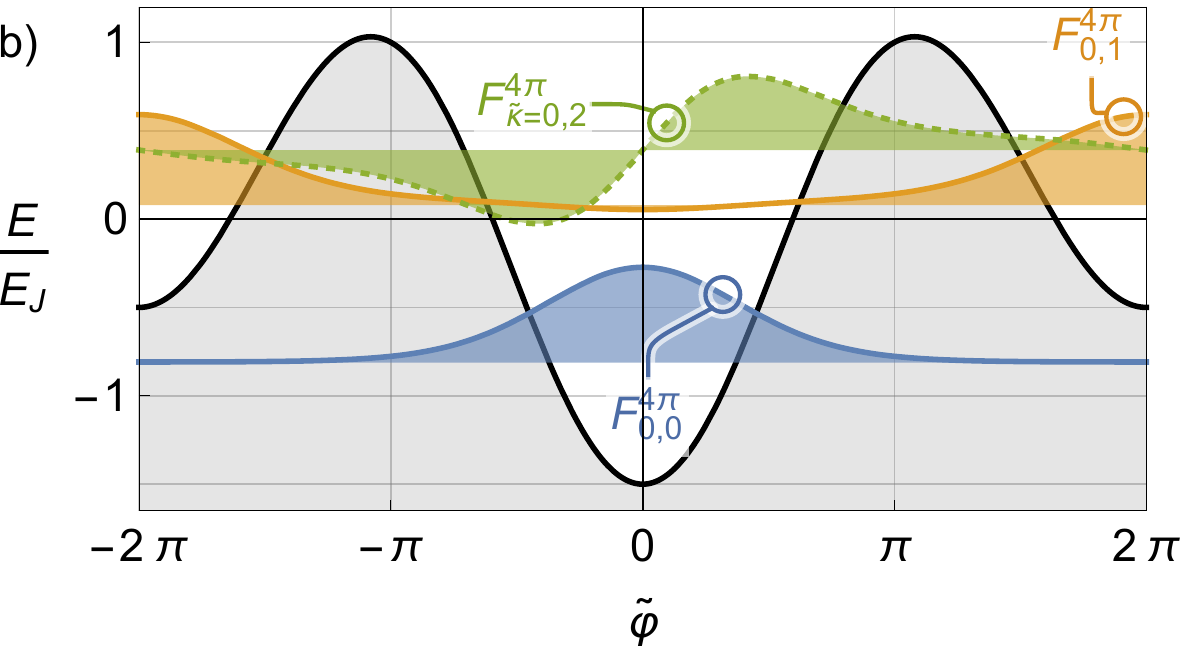}
\includegraphics[width=7.5cm]{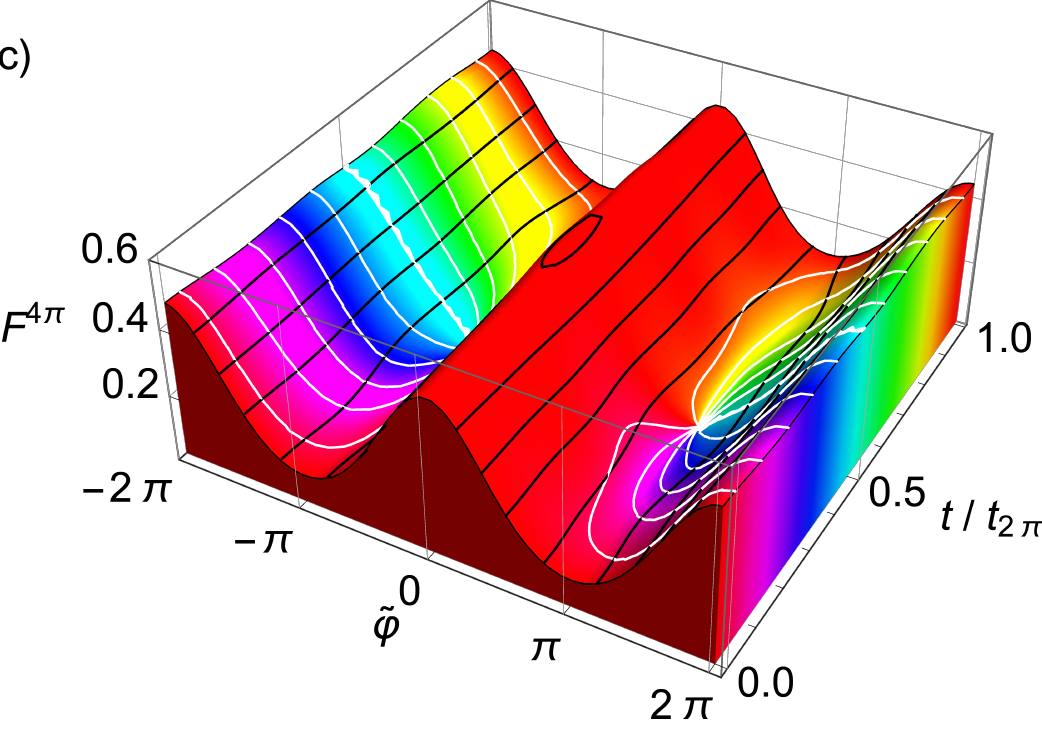}
\raisebox{0.5cm}{\includegraphics[width=0.9cm]{shuntWFlegend.pdf}}
\caption{({\sf a})  The lowest band wavefunctions, $b=0$, for the unshunted transmon ground state and the Brillouin zone edge (respectively: $\kappa=0$, solid-red; $\kappa=\nicehalfs$, dashed-blue), shown relative to the cosine potential in the extended Zak basis, \cref{4piphasebasis}.  The system is initialised in the ground state 
of the transmon Hamiltonian, $\ket{\psi^{\sf tmon}_{\kappa=0,0}}$ (solid-red). Parameters are  $E_C/E_J=1$ and $E_{\sf 4\pi}=0$.  \mbox{({\sf b})} The lowest three eigenstates of the  shunted system with $\tilde \kappa=0$, in the shunted $4\pi$-periodic potential with  $E_{\sf 4\pi}/E_J=1/2$. 
The initial wavefunction, $F^{\sf tmon}_{0,0}$,  has large and approximately equal overlap with the ground  and  first excited wavefunctions   of the $4\pi$-periodic system, $F^{\sf 4\pi}_{\tilde \kappa=0,b=0}$ and $F^{\sf 4\pi}_{0,1}$.  
({\sf c}) The system evolves from the  $2\pi$-periodic initial state $F^{\sf tmon}_{\kappa=0,0}$, through to a state at $t=t_{2\pi}/2$ that has very large overlap with the $4\pi$-periodic $F^{\sf tmon}_{\kappa=\nicehalfs,0}$ pictured in the top panel (dashed-blue).  At $t=t_{2\pi}$, the system returns to a state close to the initial state.  The scale bar shows the complex phase of the wavefunction.  
} 
\label{4pievolution}
\end{center}
\end{figure}

There are several solution strategies for this problem. Using  \cref{phasebasis}, it is straightforward to check that \mbox{$\cos(\hat\phi/2)\ket{k,\varphi}=\cos(\varphi/2)\ket{k+\nicehalfs,\varphi}$}, {and recalling that $k\in(-\nicehalfs,\nicehalfs]$  compact, so we identify $k+\nicehalfs>1$ with $k-\nicehalfs$. Thus, we have}
\begin{equation}
\cos(\hat\phi/2)=\!\int_{-\nicehalfs}^\nicehalfs \!\!\!\!dk \!\int_{-\pi}^\pi \!\!\!d\varphi \cos(\varphi/2){\ket{k+\nicehalfs,\varphi}}{\bra{k,\varphi}},\label{cosphi2}
\end{equation}
which couples states that differ in $k$ by $\nicehalfs$.  In \cref{app:4pizak} we describe a solution using this decomposition of $\cos(\hat\phi/2)$.

\begin{figure}
\begin{center}
\includegraphics[width=8cm]{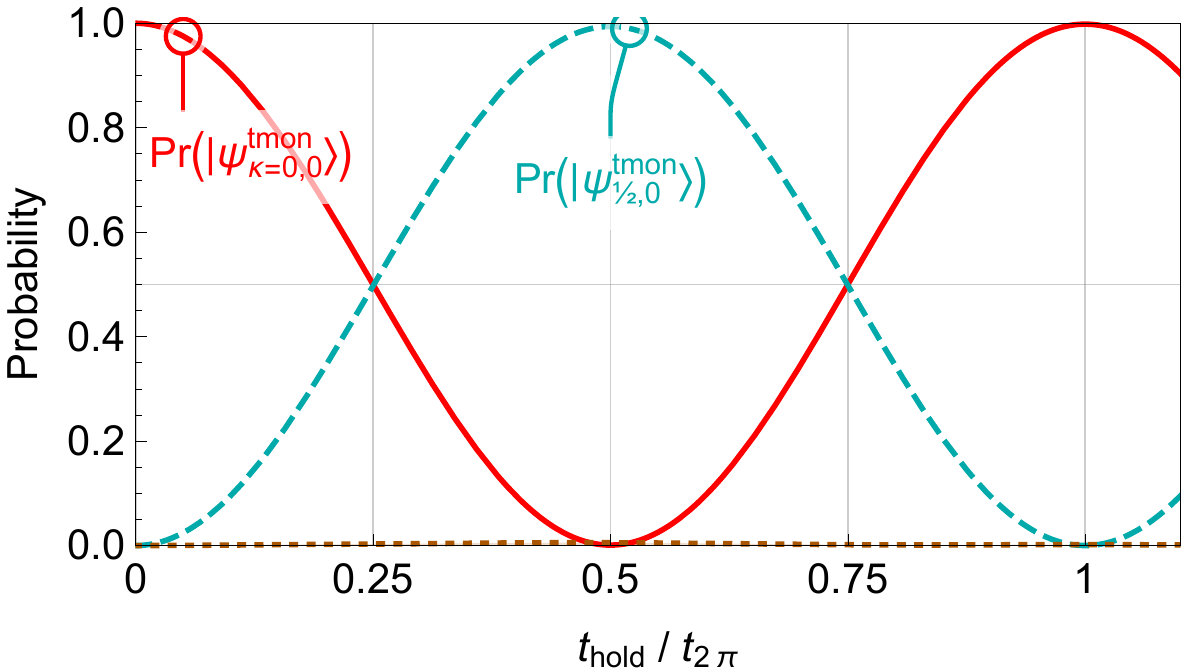}
\caption{Time-evolution of the return probability of a transmon initialised in the state $ \ket{\psi^{\sf tmon}_{0,0}}$ and then shunted by a the $4\pi$-periodic element,  (solid-red).  Also shown is the probability of the state $ \ket{\psi^{\sf tmon}_{\nicehalfs,0}}$  localised near the Brillouin zone boundary (dashed-blue), along with the residual probability in other states (dotted) which is less than 1\% for the chosen parameters,  $E_C/E_J=1$ and  $E_{\sf 4\pi}/E_J=1/2$.} \label{4pioverlap}
\end{center}
\end{figure}

 However, a more natural approach  is to note that because the $\hat\phi$-dependent potential in \cref{shuntedTransmonHamiltonian4pi} is $4\pi$-periodic for all values of $E_{\sf 4\pi}$, we can define a new $4\pi$-periodic Zak basis that accommodates this reduced symmetry.  In particular, we define the orthonormal $4\pi$-Zak basis
\begin{align}
    \ket{\tilde k,\tilde \varphi } &= 2\, {\sum}_{j\in\mathbb{Z}} \,e^{i 4\pi j \tilde k }\!\ket{\tilde \varphi - 4\pi j }_{\!\bar\phi},\label{4piphasebasis}\\
   & =\tfrac12\tfrac{e^{i \tilde k\tilde  \varphi}}{\sqrt{2\pi}}\,{\sum}_{j\in\mathbb{Z}}\,e^{-i j \tilde \varphi /2} \!\ket{ j/2-\tilde k }_{\!\bar n}.     \label{4pichargebasis}
\end{align}
where  $\tilde k\in(-1/4,1/4]$ and  $\tilde \varphi\in(-2\pi,2\pi]$.  
 The boundary conditions in this basis are
\begin{align}
 \psi( - 1/4, \tilde \varphi ) &= \psi ( 1/4, \tilde \varphi), \label{bc14pi}
\\
 \psi (\tilde k, - 2\pi ) &= e^{4\pi \tilde k i} \psi (\tilde k, 2\pi),\label{bc24pi}
\end{align}
and the system eigenstates take the form
\begin{equation}
\langle{\tilde k,\tilde \varphi}\ket{\psi^{\sf 4\pi}_{\tilde \kappa,b}} 
\equiv \psi^{\sf 4\pi}_{\tilde \kappa,b} (\tilde k,\tilde \varphi)=\delta(\tilde k-\tilde \kappa)F^{\sf 4\pi}_{\tilde \kappa,b}(\tilde \varphi),\label{eqn:deltaF4pi}
\end{equation}
where $b\in\mathbb{N}$ is a new band index.  In this basis, the eigenstates satisfy the eigenvalue problem
\begin{equation}
-E_C  \tfrac{\partial^2 \psi^{\sf 4\pi}_{\tilde \kappa,b} }{\partial \tilde \varphi^2}-\!\big(E_J\cos(\tilde \varphi)+E_{\sf 4\pi}\cos(\tilde \varphi/2)\big)\psi^{\sf 4\pi}_{\tilde \kappa,b}\!=\!E^{\sf 4\pi}_{\tilde \kappa,b} \psi^{\sf 4\pi}_{\tilde \kappa,b},\label{shuntedTransmonHamiltonian4piODE}
\end{equation}
where $E^{\sf 4\pi}_{\tilde \kappa,b}$  is the eigenenergy.  The first three  eigenfunctions at the centre of the  Brillouin zone, $F^{\sf 4\pi}_{\tilde \kappa=0,b}(\tilde \varphi)$,  of the $4\pi$-shunted transmon are shown in the middle panel of \cref{4pievolution}.

We now describe the time evolution of the system subject to a transiently shunted $4\pi$-periodic element. 
The wavefunction for the initial transmon ground state, $\ket{\psi^{\sf tmon}_{0,0}}$, is plotted in  the extended Zak basis in the top panel of \cref{4pievolution} (solid-red).  
  As in the previous section, we decompose the initial state into the eigenmodes of the shunted system, $\ket{\psi^{\sf 4\pi}_{\tilde \kappa,b}}$, and evolve in the usual way.  
Since  \mbox{$\tilde k$} is conserved,  the initial state $\ket{\psi^{\sf tmon}_{0,0}}$  only has support on modes with $\tilde \kappa=0$, i.e.\ those at the centre of the Brillouin zone in the $4\pi$-Zak basis.  Of these states, the lowest two have large overlap with the initial state, and so dominate the evolution.

    For  the numerical results here, we assume that $E_C/E_J=1$ and $E_{\sf 4\pi}/E_J=1/2$.  (This choice of parameters makes the $4\pi$-periodic potential in \cref{shuntedTransmonHamiltonian4pi} qualitatively comparable to the aperiodic potential in \cref{shuntedTransmonHamiltonian1}: the two potential functions have a global minimum at $\varphi=0$,  local minima at or near $\varphi=\pm 2\pi$, and with a potential difference between these local minima and the global minimum $\approx E_J$).  
We define the time for a coherent oscillation between the lowest and first excited states at the centre of the extended Brillouin zone (i.e.\ at 
$\tilde \kappa=0$), $t_{2\pi}=2\pi/(E^{\sf 4\pi}_{\tilde 0,1}-E^{\sf 4\pi}_{\tilde 0,0})$.  For the parameter choices in \cref{4pievolution},  $ t_{2\pi}\approx 7.04/E_J$. 

The time evolution is represented in the bottom panel of \cref{4pievolution}, starting with the initial state with $2\pi$-periodicity at $t=0$.  The system evolves to a $4\pi$-periodic state at $t_{2\pi}/2$ with a positive (red) anti-node at $\tilde\varphi=0$, and a negative (cyan)  anti-node at $\tilde\varphi=\pm2\pi$.  This  is qualitatively very similar to  $\ket{\psi^{\sf tmon}_{\nicehalfs,0}}$, as shown in the top panel of \cref{4pievolution} (dashed-blue).  At $t=t_{2\pi}$, the system returns very close to the initial state.

The probability for  the time-evolved state to be in either the initial state, $\ket{\psi^{\sf tmon}_{0,0}}$, or a state at the Brillouin zone boundary,  $\ket{\psi^{\sf tmon}_{\nicehalfs,0}}$, is shown in \cref{4pioverlap}.   The evolution occurs almost entirely in the space spanned by these two states, so that nearly perfect coherent oscillation between the initial and intermediate state is observed.  The residual probability to evolve out of this two-state subspace, shown as a dotted curve, is less than 1\% at all $t$.  Since the state evolution is unitary, the very-nearly equal probabilities for the two states at $t=t_{2\pi}/4$ and at $t=3t_{2\pi}/4$ indicate that the system is very close to the state \mbox{$(\ket{\psi^{\sf tmon}_{0,0}}+e^{i\theta}\ket{\psi^{\sf tmon}_{\nicehalfs,0}})/\sqrt{2}$} at these times.

In summary,  a $4\pi$-periodic element  switched transiently across a $2\pi$-periodic  transmon device yields evolution in a 2-dimensional (qubit-like) subspace spanned by $\{\ket{\psi^{\sf tmon}_{0,0}},\ket{\psi^{\sf tmon}_{\nicehalfs,0}}\}$.  This is closely analogous to the result that a conventional $2\pi$-periodic Josephson junction switched transiently across a $\pi$-periodic device (i.e.\ the circuit for the ${0{\text -}\pi}$ qubit) yields a computational gate in the ${0{\text -}\pi}$ encoded subspace \cite{Brooks13,kitaev2006protected}.  {Comparisons between the ${0{\text -}\pi}$ qubit and the qubit proposed here will be given in more detail in the next section. }

\section{Discussion}\label{sec:discussion}

The description of  the complete spectrum of superconducting  devices in terms of band structure, which is a  consequence of Bloch's theorem,  was identified by  \citet{likharev1985theory}.  However, it is an atypical approach to theoretical modelling of such systems,  and it leads to somewhat surprising conclusions.  

There are several natural questions that arise from the analysis above: 
\begin{inparaenum}
\item Given that the Bloch spectrum is a continuum, why does experimental spectroscopy of transmons yield apparently discrete spectra?
\item Given that $\ket{\psi^{\sf tmon}_{0,0}}$ is the unique ground state, what is the lifetime of another state  $\ket{\psi^{\sf tmon}_{\kappa,0}}$, in the lowest Bloch band?  
\item How do we interpret superpositions of Bloch eigenstates?
 \item What are the coherence properties of such superpositions?  
 \end{inparaenum}
Here we address these questions.

Firstly, our analysis shows that transmons and CPBs do not have discrete spectra, and so should not be thought of as `artificial atoms'.  Instead their spectrum is arranged in bands, with densely packed eigenstates. 
 However, intra-band transitions can only occur when the device is subject to symmetry-breaking perturbations, and so cannot be induced by external driving or other perturbations that respect the dynamical $2\pi$-phase periodicity in the Hamiltonian. {An example of such symmetry-breaking perturbations is a shunted inductor  that we have presented in \cref{sec:ind} or a shunted $4 \pi$-periodic junction as in \cref{sec:4pi}. Also, quasiparticle tunnelling introduces an environmental term $\propto \sin(\hat \phi/2)$ \cite{Catelani11}, which breaks the symmetry of concern.}  It follows that spectroscopic measurements will yield discrete  spectra corresponding to direct (i.e.\ $\kappa$-conserving) inter-band transitions, as observed in numerous experiments.

Secondly,  any perturbation that preserves the dynamical $2\pi$-phase periodicity cannot couple states $\ket{\psi^{\sf tmon}_{\kappa,b}}$ and $\ket{\psi^{\sf tmon}_{\kappa',b'}}$ with different values of the Bloch wavenumber, i.e.\ $\kappa\neq\kappa'$ \footnote{That is, the quasicharge is a conserved quantum number under symmetry preserving perturbations.}.  It follows that any state, $\ket{\psi^{\sf tmon}_{\kappa',0}}$, in the lowest band cannot relax to the true ground state $\ket{\psi^{\sf tmon}_{0,0}}$, unless the symmetry is broken.

Thirdly, our analysis indicates that  superpositions of eigenstates with different wavenumbers within the same band are also possible states of the system.  In particular, the state 
$
\ket{\Psi_b}=\alpha\ket{\psi^{\sf tmon}_{\kappa,b}}+\beta\ket{\psi^{\sf tmon}_{\kappa'\!,b}}
$,  introduced in \cref{superposition}, should be physically admissible (just as superpositions of different band indices, $b$, are currently used to define qubit states of transmons and CPBs).  A physical interpretation can be given to certain such superpositions.  For example, the state $\ket{\psi^{\sf 4\pi}_{\tilde \kappa=0,0}} \approx\ket{\psi^{\sf tmon}_{0,0}}+\ket{\psi^{\sf tmon}_{\nicehalfs,0}}$ is the ground state of the $4\pi$-periodic shunted Hamiltonian \cref{shuntedTransmonHamiltonian4pi},  pictured as the lowest mode in the middle panel of \cref{4pievolution}, so is a state with $4\pi$-periodicity in the superconducting phase.  In the charge basis, the state will have $1e$ charge periodicity.  

Finally, as shown in Ref. \cite{le2019doubly}, dephasing of a coherent superposition of energy eigenstates due to perturbative, stochastic charge noise depends on the gradient of the band, $\Gamma\propto |\partial_\kappa E_{\kappa,b}|^2$. Since the energy bands are extremal at both the centre and the edge of the Brillouin zone (i.e.\ the gradient vanishes), we anticipate that the coherent superposition 
$
\alpha \ket{\psi^{\sf tmon}_{0,0}}+\beta \ket{\psi^{\sf tmon}_{\nicehalfs,0}}\nonumber
$ in \cref{superposition} should have extremely long coherence times in the presence of perturbative charge noise.  Deep in the transmon regime, where the lowest band is almost flat, this result will hold even for  charge fluctuations of order $|\delta n_x|\lesssim1/2$.

\subsection{A possible qubit encoding}

As a consequence of these facts, we expect that any state of the form $\ket{\Psi_{b=0}}$ will have extremely long  $T_1$ and $T_2$ times. Further, the two states $\ket{\psi^{\sf tmon}_{0,0}}$ and $\ket{\psi^{\sf tmon}_{\nicehalfs,0}}$, pictured in the lowest band of \cref{bandstructure}, are approximations to the comb-like states proposed for encoding a qubit in an oscillator \cite{Gottesman01}, and so may inherit some of the robustness of such comb states as qubit encodings. 

We therefore speculate that  the subspace spanned by these states in the lowest band provides an attractive home for a qubit, with computational states \mbox{$\ket{0}= \ket{\psi^{\sf tmon}_{0,0}}$} and \mbox{$\ket{1}= \ket{\psi^{\sf tmon}_{\nicehalfs,0}}$}.
{This choice  contrasts with the conventional transmon qubit encoding  in which the logical ``1'' state is encoded in the first excited band, i.e.\ \mbox{$\ket{\bar 1}\equiv \ket{\psi^{\sf tmon}_{0,1}}$}.}  We outline here possible single-qubit gates and spectroscopic readout.  

 High-quality Pauli-$X$ rotations in this encoding could be generated by transiently switching a $4\pi$-periodic element across the circuit, for a well controlled hold time, as shown in \cref{4pioverlap}.  If $E_J\gg E_C$ (the transmon regime), the computational states are nearly degenerate, but this near-degeneracy can be controllably lifted by tuning $E_J$ dynamically into the regime where $E_J\sim E_C$, which will result in a Pauli-$Z$ rotation in the computational space.  Experimentally, control over $E_J$ may be achieved by replacing the junction in \cref{fig:transmon} with a flux-tunable SQUID.  We note these operations would not be natively robust against fluctuations in the control parameters, however some of the ideas in Ref. \cite{Brooks13} for constructing robust gates for the $0{\text -}\pi$ qubit might be adapted to the encoding suggested   here.

Readout in the computational basis can be achieved using spectroscopic methods: \cref{bandstructure} shows that the transition energy from the computational states $\ket{0}$ and $\ket{1}$  to the $b=1$ band will have different energies depending on the computational state.  A reflectometry experiment with a probe field tuned to the $\ket{\psi^{\sf tmon}_{\kappa=0,b=0}}\leftrightarrow\ket{\psi^{\sf tmon}_{\kappa=0,b=1}}$ transition will show strong response if the system were in the state \mbox{$\ket{0}= \ket{\psi^{\sf tmon}_{0,0}}$}, and very little response if the system were in the state \mbox{$\ket{1}= \ket{\psi^{\sf tmon}_{\nicehalfs,0}}$}.  Thus, the observation (or not) of significant reflected power would project the system into the corresponding computational state, facilitating  computational readout.

Finally, the fact that the proposed qubit states have comb-like representations akin to the `GKP' encoding of a qubit in an oscillator  \cite{Gottesman01}, suggest that active error correction against charge noise could be implemented based on the structure of that code.  To do so would require (i) that the comb peaks to be narrow, i.e.$\sqrt{E_C/E_J}\ll1$, where the device  is in the transmon regime,  (ii) a mechanism to nondestructively measure the displacement of the modular phase arising from charge noise, and (iii) an actuator to act back on the system to correct for drifts in the modular phase.  A protocol to achieve (ii) using spectroscopically resolved transitions to the first excited band in a related system was briefly mentioned in Ref. \cite{le2019doubly}, however a complete analysis of such a protocol in the presence of realistic noise models will be the subject of future research.

\subsubsection*{{Comparisons with the 0-$\pi$ qubit}}

\begin{table}[t]
\caption{{Comparisons between the 0-$\pi$ qubit and the non-compact transmon/Cooper-pair box. The qubit encoding proposed here is based on a transmon/Cooper-pair box using a conventional $2\pi$ Josephson junction, but using states only in the lowest band.  To perform gates a device with a lower symmetry is required.  The 0-$\pi$ qubit is based on a high symmetry junction ($\pi$ periodic), but uses a conventional $2\pi$ junction for gates.}}

  \begin{tabular}{|P{4cm}|P{2cm}|P{2cm}|}
    \hline
     & Qubit potential periodicity         & Gate potential periodicity \\ \hline
  The 0-$\pi$ qubit  \cite{kitaev2006protected,Brooks13}                 & $\pi$ & $2\pi$   \\ \hline
   The non-compact transmon/Cooper-pair box                     &  $2\pi$  & $4\pi$   \\ \hline
  \end{tabular}
  \label{table:compare}
  \end{table}

{Table \ref{table:compare} shows comparisons between the 0-$\pi$ qubit \cite{kitaev2006protected,Brooks13} and the non-compact transmon/CPB proposed here. The circuit element required to construct the 0-$\pi$ qubit is an effective $\pi$-periodic Josephson junction ladder \cite{Brooks13,Dempster14}. The relevant gate operations on the 0-$\pi$ qubit need an ordinary $2\pi$ Josephson junction. The 0-$\pi$ qubit has been studied in various theoretical works \cite{Brooks13,kitaev2006protected,Paolo19,Dempster14,Groszkowski2018} and recently has been realised in experiment \cite{gyenis2019}.}

{By contrast, the circuit for the non-compact transmon/CPB qubit has been well developed experimentally \cite{Nakamura99,Schreier08,Arute19}, but gate operations, as shown in \cref{sec:4pi}, would be ideally performed by a $4\pi$-periodic Josephson junction. This element is a current experimental challenge and is being investigated in the context of the Majorana qubit \cite{Ginossar:2014aa,PhysRevX.6.031016,Laroche:2019aa,Mota19}. A proposal for a $4\pi$-periodic element has been put forward by Bell \textit{et al.} \cite{Bell16}. }

\subsection{Experimental considerations}

To our knowledge, superpositions such as  \cref{superposition} have not been  experimentally reported. One reason for this  is that it is technically challenging to prepare and measure such a state.  To do so, it is necessary to controllably break the dynamical $2\pi$-phase periodicity  of the JJ element.  As described in \cref{sec:symmetrybreaking}, a transiently switched inductive element achieves this.

This discussion suggests specific experiments that might be developed to demonstrate the possibility of producing and measuring superpositions of Bloch eigenstates.  For example, a device that is intermediate between a transmon and a CPB, with $E_J\lesssim E_C$ will have spectrally resolvable transitions at different values of $\kappa$.  Repeatedly preparing the ground state of such a system, then  shunting a symmetry breaking inductive element across it for different hold times, $\thold$, and then spectroscopically measuring the state via reflectometry should map out coherent oscillations as shown in \cref{indoverlap} and \cref{4pioverlap}.  This would constitute  evidence for the non-compact description of the \mbox{charge and flux discussed here.}

One of the key experimental hurdles to such a demonstration  is the requirement of a fast, dissipation-free switch.  This element, depicted in \cref{fig:transmon}, is required to transiently shunt the inductive element across the circuit.  One common approach to building switched devices in superconducting systems is to use a flux-tunable SQUID, in which $E_J(\Phi_x)$ can be switched from large to small values quickly.  For small fluctuations of the modular phase, $\varphi$, this can be treated as a modulation of a linearised inductance.  However SQUIDs are natively $2\pi$-periodic devices, so they are not suitable for switching devices that break this symmetry.  Instead, the switch should ideally present infinite (linear) impedance when it is open, and zero impedance when closed, so that the circuit sees only the additional load due to the switched inductive shunt element.  One approach to this is to use nano-mechanical metallic plates that can be forced in-or-out of contact with one another using electrostatic control fields \cite{Czaplewski_2009} to build a fast, nano-scale mechanical switch.  Another approach is to build a hybrid super-semiconducting device, using e.g.\ a gated superconducting-normal-superconducting junction \cite{PhysRevLett.81.1682}, or  semi-superconducting hybrids \cite{Bustarret:2006aa,ridderbos2019hard,gill2016hybrid}  which might enable a controllable superconducting switch  \cite{Dienst:2011aa,Thierschmann18}.

In the short term, the general principles discussed here could be demonstrated with a switched linear inductor.  However   comparison of the solid-red curves in \cref{indoverlap} and \cref{4pioverlap} shows that for producing well-controlled superpositions of Bloch states, the (phenomenological) $4\pi$-periodic nonlinear inductive element is likely to result in better control of the system. Thus a secondary hurdle in future experiments is the development of robust $4\pi$-periodic elements. While we have adopted a purely phenomenological description of such an element, realistic implementations are the subject of  recent theoretical and experimental attention \cite{PhysRevB.77.184506,PhysRevB.79.161408,Ginossar:2014aa,PhysRevX.6.031016,Laroche:2019aa,Mota19}.  Understanding the performance of the switched system using  realistic models  for practical  $4\pi$-periodic elements will be the subject of future research.

\section{Conclusions}

We have revisited the question of the complete Hilbert space required to describe superconducting circuits such as transmons and CPBs.  Lifting the assumption that charge is discrete (or flux is compact) yields a Bloch band structure arising from the dynamical $2\pi$-periodicity in the Josephson junction phase.  We have shown that Bloch eigenstates can be coupled using elements that break the dynamical $2\pi$-periodicity, including linear and nonlinear inductive elements transiently switched across the circuit.  The resulting evolution generates states that are superpositions  of the Bloch eigenstates.  We speculate that certain eigenmodes in the lowest Bloch band would be naturally stable qubit states  with extremely long lived population life-times, and we anticipate that they will have native robustness against charge noise.  Finally, we have suggested a class of experiments that could test the results in this paper.

\begin{acknowledgments}
This research was supported by the Australian Research Council Centres of Excellence for Engineered Quantum Systems (EQUS, CE170100009), and Future Low-Energy Electronics Technologies (FLEET, CE170100039). 
TMS acknowledges visitor support from the Pauli Center for Theoretical Studies, ETH Zurich.  We thank Clemens M\"uller, Samuel Wilkinson, Victor Albert, and Arne Grimsmo for useful comments. 
\end{acknowledgments}

\appendix

\begin{figure}[t]
\begin{center}
\includegraphics[width=4.25cm]{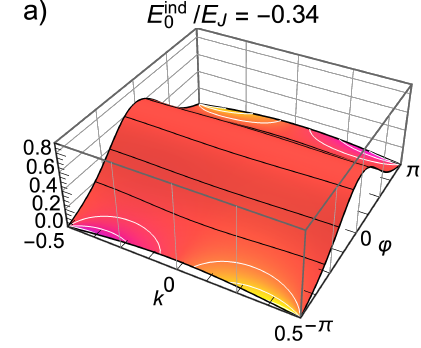}
\includegraphics[width=4.25cm]{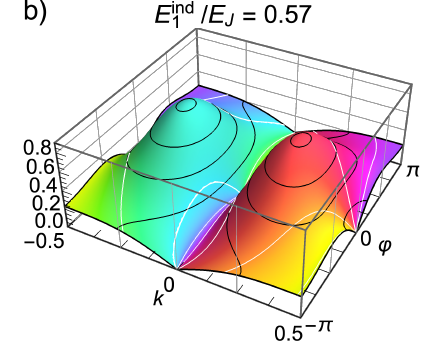}
\includegraphics[width=4.25cm]{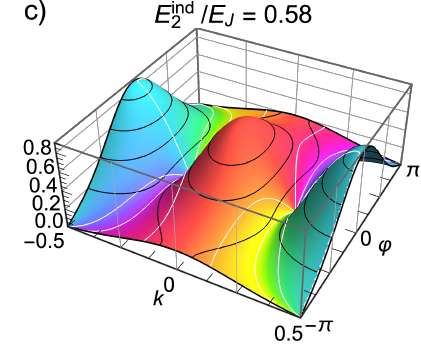}
\includegraphics[width=4.25cm]{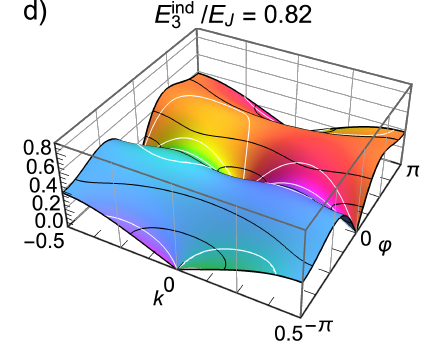}
\includegraphics[width=4.25cm]{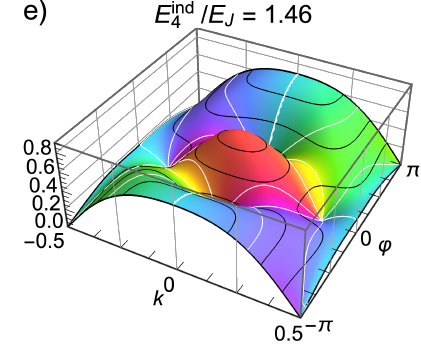}
\includegraphics[width=4.25cm]{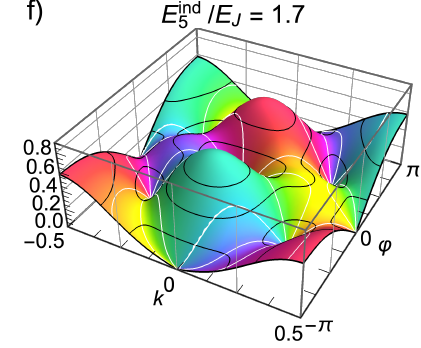}
\includegraphics[width=4.25cm]{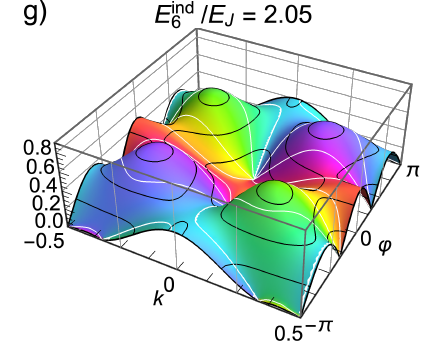}
\hfill
\mbox{{\includegraphics[width=0.9cm]{shuntWFlegend.pdf}
\hphantom{2cm}}}
\caption{(a-g) The first 7 complex-valued eigenmodes of the inductively shunted transmon {(i.e. the fluxonium)}, in the Zak basis, $\psi_j^{\sf ind}({k,\varphi})$.  The  corresponding, calculated eigenenergies $E^{\sf ind}_j/E_J$ are shown above each plot, for the parameter values $E_C/E_J=1$ and $E_L/E_J=1/(2\pi)^2$.  For the time-evolution shown in \cref{fig:timeEvolution}, modes $j=0$ and 2 have large overlap with the initial state (shown at the left of \cref{fig:timeEvolution}).  Colours indicate the phase of the wavefunction at each point, as specified in the legend (bottom right).
} 
\label{shuntedWF}
\end{center}
\end{figure}

\section{{Mode functions for the inductively shunted transmon (the fluxonium) in the Zak Basis}}
\label{app:indeigenmodes}

In this appendix we show the first 7 eigenfunctions, which are labelled by the corresponding eigenenergies $E^{\sf ind}_j$ of the inductively shunted transmon, {i.e. the fluxonium}, arrayed in \cref{shuntedWF}, in the Zak basis, for the parameter values $E_C/E_J=1$ and $E_L/E_J=1/(2\pi)^2$. {The eigenfunctions are found by solving the Schr\"odinger equation \cref{shuntedTransmonHamiltonian}
subject to the two boundary conditions defined in Eqs. \eqref{bc1} and \eqref{bc2}.}

 There is an approximate degeneracy between $E^{\sf ind}_1$ and $E^{\sf ind}_2$, for which the corresponding eigenmodes are  localised around $\varphi=0$.  The eigenfunctions exhibit qualitative features of the boundary conditions \cref{bc1,bc2}. {Notably, they are generally delocalised in the modular charge $k$. This is different from the transiently inductor-shunted transmon/CPB logical states $\ket{\psi_{0,0}^{\sf tmon}}$ and $\ket{\psi_{\nicehalfs,0}^{\sf tmon}}$, which as shown in the first and middle snapshots of \cref {fig:timeEvolution} are localised around $k=0$ and $k = 1/2$, respectively. The eigenfunctions, additionally, are symmetric about the modular phase $\varphi$, which is similar to the case of $\ket{\psi_{0,0}^{\sf tmon}}$ and $\ket{\psi_{\nicehalfs,0}^{\sf tmon}}$}.
 


 In the limit of large $j$, we expect the modes to be dominated by the harmonic terms, so that \mbox{$\lim_{j\rightarrow\infty}E^{\sf ind}_{j+1}-E^{\sf ind}_j=E_{\rm HO}=2\sqrt{E_C E_L}$}.  For the choice of parameters here, $E_{\rm HO}=E_J/\pi$, and we confirm that this  holds numerically for  $j\gtrsim 30$.

\begin{figure}[t!]
\begin{center}
\includegraphics[width=8cm]{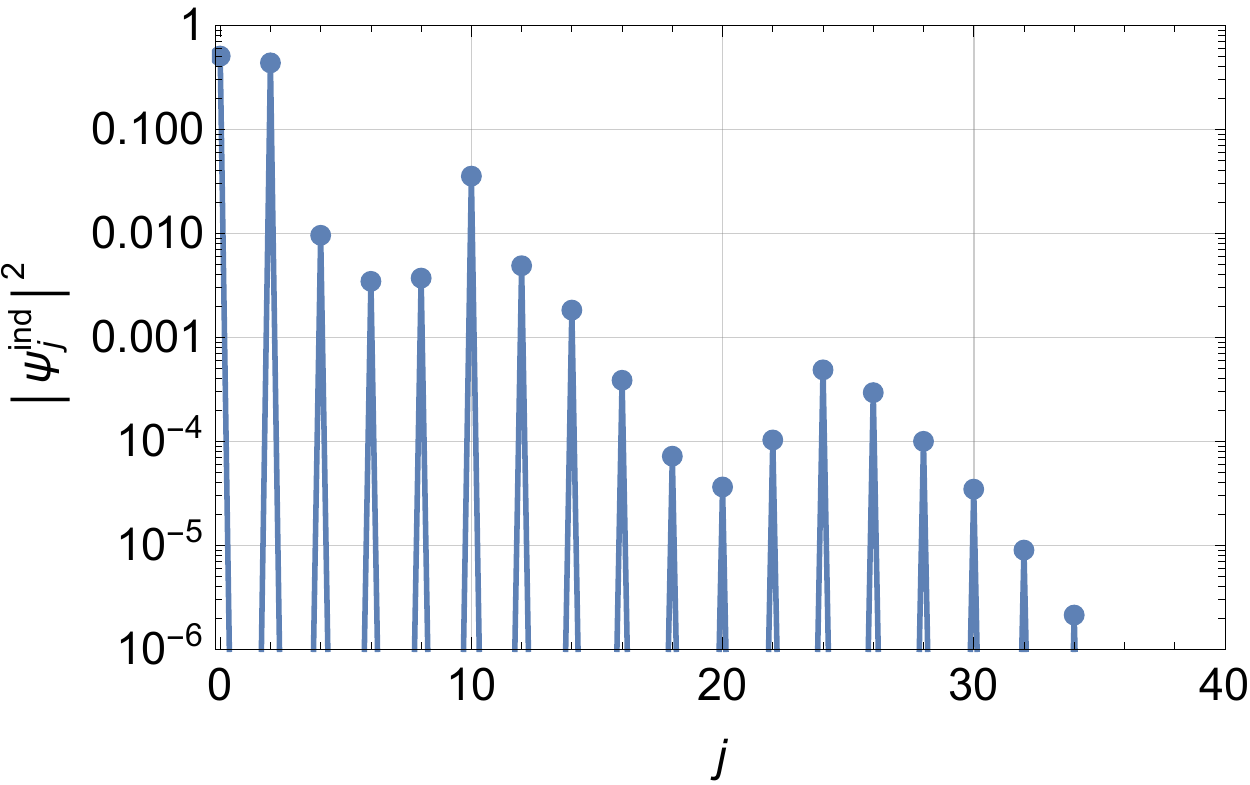}
\caption{Probabilities for each mode in the inductively shunted transmon used to calculate the time evolution in \cref{fig:timeEvolution}.  Odd-numbered modes have zero amplitude.  The initial state has overwhelming support on the ground and second excited state, making the evolution approximately that of a two-level system.} \label{coeffs}
\end{center}
\end{figure}

For these parameter values, and the initial state \cref{eqn:indinitialstate}, the mode probabilities are plotted in \cref{coeffs}, $ |p_j^{\sf ind}|^2=|\langle \psi_j^{\sf ind} \ket{\psi^{\sf tmon}_{0,0}}_{\!\Delta}|^2$.  The initial state has no support on odd numbered eigenmodes, and the majority of the probability density has support on the ground and second excited states, accounting for about 95\% of the total probability density.  Thus, the system could be approximated by a two level system, consisting of these two modes.  For numerical simulations of temporal evolution, we retain the first 100 such modes, which accounts for $>1-10^{-5}$ of the total probability. 

\section{Solution for $4\pi$-periodic Hamiltonian in the Zak basis}
\label{app:4pizak}

The $4\pi$-periodic element   conserves \mbox{$\tilde k\equiv (k \mod \nicehalfs)\in(-1/4,1/4]$}, and couples states separated by $\delta k=\nicehalfs$.  The time evolved state will therefore be of the form
\begin{equation}
\ket{\psi^{\sf 4\pi}(t)}=\ket{\psi^{\sf 4\pi}_{\kappa=0}(t)}+\ket{\psi^{\sf 4\pi}_{\nicehalfs}(t)},\label{timeevpsi}
\end{equation}
where the two (unnormalised) states on the RHS of \cref{timeevpsi} are localised around $\kappa=0$ and $\nicehalfs$ respectively.  In the Zak basis the corresponding  wavefunctions  $f^{\sf 4\pi}_{\kappa}(t;\varphi)$ are defined implicitly by
\begin{equation}
\psi^{\sf 4\pi}_{\kappa} (t;k, \varphi)\equiv\langle{k,\varphi}\ket{\psi^{\sf 4\pi}_{\kappa}(t)}=\delta(k-\kappa)f^{\sf 4\pi}_{\kappa}(t;\varphi).\label{eqn:4pif}
\end{equation}
The wavefunctions  
satisfy the coupled Schr\"odinger equations (for brevity, we have suppressed the argument to $f^{\sf 4\pi}_{\kappa}(t;\varphi)$)
\begin{align}
i\tfrac{\partial}{\partial t} f^{\sf 4\pi}_{0}&\!=\!-\!E_C \tfrac{\partial^2}{\partial \varphi^2}  f^{\sf 4\pi}_{0}\!-\!E_J \cos(\varphi)f^{\sf 4\pi}_{0}\!-\!E_{\sf 4\pi} \cos(\tfrac{\varphi}{2})f^{\sf 4\pi}_{\nicehalfs},\nonumber\\
i\tfrac{\partial}{\partial t} f^{\sf 4\pi}_{\nicehalfs}&\!=\!-\!E_C \tfrac{\partial^2}{\partial \varphi^2}  f^{\sf 4\pi}_{\nicehalfs}\!-\!E_J \cos(\varphi)f^{\sf 4\pi}_{\nicehalfs}\!-\!E_{\sf 4\pi} \cos(\tfrac{\varphi}{2})f^{\sf 4\pi}_{0},\label{SEf}
\end{align}
and boundary conditions
$
f^{\sf 4\pi}_{\kappa}(t;-\pi)=e^{2i\pi\kappa}f^{\sf 4\pi}_{\kappa}(t;\pi),
$
consistent with the boundary condition \cref{bc2}.

In the usual way, the linear system, \cref{SEf}, can be turned into an eigenvalue problem 
that is characterised by pairs of functions $\{f^{\sf 4\pi}_{0}(\varphi),f^{\sf 4\pi}_{\nicehalfs}(\varphi)\}_j$, associated to each eigenenergy $E_j$. We have confirmed that this approach gives the same results as the method described in the main text.

\bibliography{qpsjj}


\end{document}